\documentclass[aps,prl,superscriptaddress,10pt,twocolumn]{revtex4-2}
\usepackage{bibunits} % For multiple bibliographies

\usepackage{hyperref} % Optional: For clickable links in references
\usepackage[utf8]{inputenc}
\usepackage[english]{babel}
\usepackage[T1]{fontenc}
\usepackage{lmodern}
\usepackage{amsmath,amsfonts,amssymb,amsthm,bm,times,dcolumn}
\usepackage{microtype}
\usepackage{braket}
\usepackage{gensymb} % for example \degree symbol can be used!
\usepackage{physics}
\usepackage{color}
\usepackage{soul}
\usepackage[normalem]{ulem}
\usepackage{graphicx,color}
\defaultbibliography{topo4thermo-20260717-arXiv}
\begin{document}

\title{Classifying Topology via Edge-State Pure Thermalization}
%\title{Micromaser Topology Classifier via Edge-State Pure Thermalization}
%\title{Topology determines the thermodynamic character of quantum fuel}

	\author{Fatih Ozaydin}
	%\email{mansursah@gmail.com}
	%\affiliation{Department of Electrical and Computer Engineering, Lyle School of Engineering, Southern Methodist University, Dallas, TX, USA}
	\affiliation{Institute for International Strategy and Emerging Technologies, Tokyo International University, 4-42-31 Higashi-Ikebukuro, Toshima-ku, Tokyo 170-0013, Japan}
	\affiliation{Nanoelectronics Research Center, Kosuyolu Mah., Lambaci Sok., Kosuyolu Sit., No:9E/3  Kadikoy, Istanbul, T\"urkiye}
	
	\author{Haydar Sahin}
	\affiliation{Department of Electrical and Computer Engineering, National University of Singapore, Singapore 117583, Republic of Singapore}
	\affiliation{Institute of High Performance Computing, A*STAR, Singapore 138632, Republic of Singapore}
	
	\author{Armando Perez‐Leija}
	\affiliation{Department of Electrical and Computer Engineering, Saint Louis University, St Louis, Missouri, 63103, USA}
	
	\author{Azmi Ali Altintas}
	\affiliation{Department of Physics, Faculty of Science, \.{I}stanbul University, Vezneciler, \.{I}stanbul, 34116, Türkiye}
	
	\author{Cihan Bay\i nd\i r}
	\affiliation{Engineering Faculty, \.{I}stanbul Technical University, Sar\i yer \.{I}stanbul, 34469, Türkiye }
	
	\author{\"{O}zg\"{u}r E. M\"{u}stecapl{\i}o\u{g}lu}
	\affiliation{Department of Physics, Ko\c{c} University, Sar{\i}yer, \.Istanbul, 34450, Türkiye}
	\affiliation{TÜB\.ITAK Research Institute for Fundamental Sciences (TBAE), 41470 Gebze, Türkiye}
	
	\author{\c{S}ahin Kaya \"{O}zdemir}	
	\affiliation{Department of Electrical and Computer Engineering, Saint Louis University, St Louis, Missouri, 63103, USA}
		
	\date{\today}

\begin{abstract}
Repeated-interaction machines distinguish heat-like from work-like resources through the steady states they generate, but whether topology can control this distinction remains unknown. Here we reveal the role of topology in the process by showing that topological edge states can act as pure-thermalization fuels. For an open Su--Schrieffer--Heeger chain used as the fuel source of a micromaser, edge eigenstates suppress both displacement and squeezing and drive the cavity to a Gibbs state, whereas bulk eigenstates activate coherent channels and yield thermo-mechanical operation. This edge–bulk thermodynamic dichotomy remains robust under realistic decoherence, cavity loss, bond disorder, and moderate onsite disorder. We further design a superconducting implementation in which a sixteen-site SSH eigenstate is deterministically compressed into a four-qubit fuel register. The resulting cavity response provides a transport-free classifier of topology and identifies a topology--thermodynamics link that extends beyond cavity-QED to repeated-interaction settings more generally.
\end{abstract}
	
\maketitle
Repeated-interaction architectures, such as micromasers, provide a versatile framework for investigating quantum thermodynamic processes~\cite{lamb1964theory, filipowicz1986theory}. In these architectures, a stream of ancillary states acts as ``quantum fuel'' that sequentially couples to a target quantum system. This discrete-time interaction effectively drives the target system towards a steady state wherein work-like and heat-like (i.e., thermal) resources can be clearly differentiated. This is understood from the fact that stochastic thermal ancillas lead to entropy production and Gibbs relaxation of the target system; while ancillas exhibiting coherence features—such as displacement or squeezing—enable the system to reach a steady state from which ordered energy can be extracted. That is, examining the resulting steady-state helps classify these interactions into purely thermal or thermo-mechanical processes, providing a clear path to design autonomous quantum heat engines~\cite{dag2016multiatom}. This distinction is practically realized through the binary signature generated by heat-only fuels, which not only facilitates autonomous thermal protocols but also serves as a versatile tool for characterizing arrival statistics and loss mechanisms in certain platforms such as superconducting cavities.

Simply put, quantum fuels that achieve pure thermalization offer three distinct advantages for the characterization of quantum systems: first, they operationally isolate heat transfer, allowing for unambiguous thermodynamic accounting. Second, they facilitate passive reservoir engineering by establishing a robust Gibbs reference for thermometry and state reset. And third, they serve as a critical test for foundational principles, such as the bounds of ergotropy and the nature of passive states.

Prior investigations have identified several families of quantum fuels that satisfy the pure thermalization condition, including structured dimers and multi-qubit clusters~\cite{tuncer2019work, dag2019temperature, ozaydin2024engineering}. %Similar operational partitions between the coherent (work-like) and incoherent (heat-like) contributions of quantum ancillas have been explored in the context of generalized Carnot limits~\cite{niedenzu2018quantum} and the entropy production of complex collision models \cite{rodrigues2019thermodynamics}, supporting the feasibility of coherence-mediated temperature tuning without active work extraction.
Related studies have also identified operational separations between the coherent (work-like) and incoherent (heat-like) contributions of quantum ancillas. 
Such partitions appear, for example, in generalized Carnot bounds for coherence-assisted thermal machines~\cite{niedenzu2018quantum} and in the entropy production of complex collision models~\cite{rodrigues2019thermodynamics}. 
These results support the feasibility of coherence-mediated temperature tuning without active work extraction.

A promising candidate for the realization of such fuels lies in the topology degree of freedom. Topological band theory predicts robust, exponentially localized edge modes protected by global symmetries, with the Su–Schrieffer–Heeger (SSH) chain serving as the canonical minimal model~\cite{su1980soliton}. In finite open chains, the nontrivial phase hosts mid-gap edge eigenstates whose localization properties are intrinsically linked to the bulk gap. While these edge modes are widely exploited for their transport robustness, their thermodynamic manifestations have primarily been explored through external probes, such as calorimetric or heat-transfer diagnostics. 
Notable examples include thermoelectric detection in Majorana nanowires~\cite{lopez2014thermoelectrical}, stroboscopic heat-current measurements of Floquet-Majorana phases~\cite{molignini2017sensing} and specific-heat characterizations of Majorana surface states~\cite{bunkov2020direct}. Related studies have also framed topology through its impact on the performance of heat engines or as an equilibrium marker~\cite{wang2020performance, fadaie2018topological, quelle2016thermodynamic, bohling2018thermoelectric, yunt2020internal, many2022nonlinear, hu2022observation, bowick2022symmetry}. 
%\newpage

%\newpage

\onecolumngrid
\begin{widetext}
	\onecolumngrid
	\onecolumngrid
	\begin{figure}[t!]
		\includegraphics[width=0.95\linewidth]{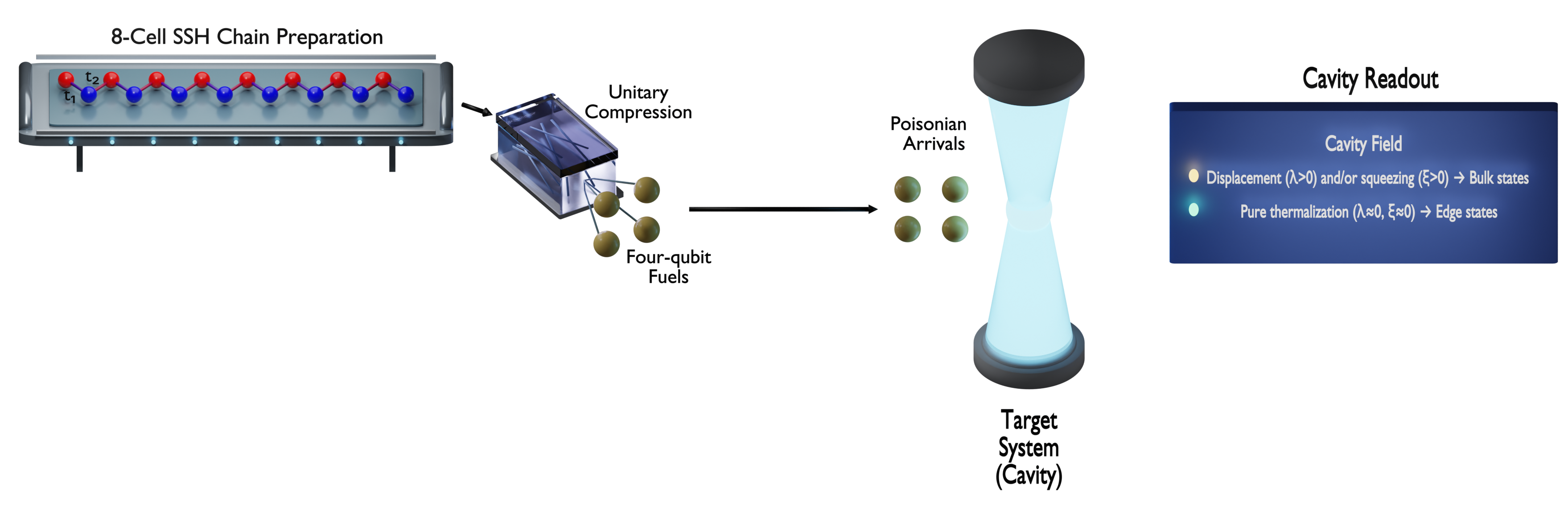}
		\caption{\textbf{Conceptual schematic of the edge–fuel micromaser.}
			An open 8-cell SSH chain is encoded on four qubits; a chosen eigenstate $|v_s\rangle$ (bulk or edge) is prepared as the \emph{fuel}, undergoes generalized amplitude damping during flight ($t_{\mathrm{tr}}$), and interacts with the target system (e.g., a single cavity mode via the Tavis–Cummings Hamiltonian) with coupling $g$ and interaction time $\tau$, at Poissonian arrival times. The cavity response operationally distinguishes the fuels: \emph{edge eigenstates} (``edge fuels'') activate only heat-exchange channels and drive a \emph{purely thermal (Gibbs)} cavity, whereas \emph{bulk eigenstates} activate coherent channels (coherent displacement and squeezing) allowing ordered energy extraction.  $t_1$ and $t_2$ represent the intra- and inter-cell hopping amplitudes.\label{fig:figconcept}}
	\end{figure} 
\end{widetext}
\twocolumngrid

 Despite extensive research into the thermal probes of topological systems, the question remains whether the symmetry-protected structure of topological states can be utilized to operationally isolate heat from work. 
%\noindent 
%Specifically, the inherent suppression of coherent drives required by ``pure-thermalization'' fuels suggests that topological edge states might offer a more robust resource than traditional dimers or multi-qubit clusters. 
Specifically, we seek to answer the question of whether the suppression of coherent drives in ``pure-thermalization'' protocols is leveraged by topological edge states to surpass the performance of traditional dimers or multi-qubit clusters as a robust resource.

In this work, we demonstrate that the topology degree of freedom, manifested as edge states, provides a robust mechanism for engineering a new class of pure-thermalization fuels. We achieve this by utilizing topological edge states as fuel in repeated interaction micromaser systems. Our findings reveal that the unique structure of topological edge states naturally prevents displacement and squeezing and drives the associated cavity maser into a purely thermal Gibbs state. In contrast, bulk states yield a complex thermo-mechanical response.

To show these effects, we implement a protocol in which an eigenstate of a finite Su--Schrieffer--Heeger (SSH) chain is first prepared and then unitarily compressed into a minimal qubit register. 
This compression step maps the single-particle lattice eigenstate onto a finite-dimensional ancilla that can be directly interfaced with a cavity mode, enabling repeated and well-controlled collision dynamics. 
In this way, the relevant topological information is preserved while the system is recast into a form suitable for thermodynamic interrogation.
This architecture enables a transport-free, binary topology classifier via quadrature readout, providing a robust signature even in the presence of disorder, loss, and decoherence. 
Crucially, the scheme is platform-agnostic, offering a general framework for thermodynamic control across diverse quantum architectures.

Concretely, the platform considered here as an example is a cavity-based repeated-interaction micromaser: 
A sequence of four-qubit atomic ensembles—each prepared in a prescribed state derived from an open eight-cell SSH chain—is injected into a lossy cavity at Poissonian arrival times. Once inside, the atoms couple to a single cavity mode via a Tavis–Cummings interaction before exiting the cavity.
The cavity quadratures then provide a direct operational readout of the fuel-induced coherent action. Within this setup, our central result is a sharp edge--vs--bulk dichotomy: fuels prepared from the two topological edge eigenstates suppress both coherent channels and drive the cavity to a purely thermal Gibbs state, whereas fuels prepared from the bulk eigenstates activate coherent displacement $(\lambda)$ and squeezing $(\xi)$, yielding thermo-mechanical response.

Our protocol follows a four-stage workflow: (i) Initialization: We prepare a single-excitation eigenstate within a physical SSH lattice. (ii) Unitary compression: This state is mapped onto a compact four-qubit register, which serves as the active ``fuel'' by preserving the original topological signature in a reduced Hilbert space. (iii) Interaction: The cavity is driven by a stream of these registers via a repeated-interaction protocol. (iv) Diagnostics: A steady-state analysis  based on monitoring the field quadratures distinguishes between coherent work extraction and pure-thermalization (heat-only) behaviour.

We represent the SSH source in the \emph{single–excitation manifold} of an open chain with eight dimers (16 sites), whose Hamiltonian $H_{\mathrm{SSH}}(t_1,t_2)$, with $t_1$ and $t_2$ representing intra- and inter-cell hoping amplitudes, supports two edge eigenstates in the topological regime ($t_1<t_2$) and fourteen bulk eigenstates across parameters.
A sharp dichotomy emerges: \emph{only the two topological edge eigenstates} drive the target system to a \emph{purely thermal} Gibbs steady state, whereas \emph{all fourteen bulk eigenstates} activate the coherent channels. The dichotomy persists under realistic imperfections within previously established parameter ranges~\cite{dag2019temperature,ozaydin2024engineering}. We refer to a fuel prepared in a topological edge eigenstate as an \emph{edge fuel}. We outline practical preparation routes for these fuels, and implementation details are given in {\color{blue}Supplementary Information, Section S1}. 
This implementation route is included as an illustrative state-preparation example and is not essential to the central thermodynamic result.%, which depends only on the properties of the prepared fuel states.
%Note that the presented superconducting protocol is intended only as an illustrative route for state preparation and transfer; the central thermodynamic conclusions of this work depend only on the resulting fuel states, not on any specific preparation method.

Because concrete implementations are compatible with current superconducting-cavity parameters, including realistic cavity loss and decoherence~\cite{tuncer2019work, ozaydin2024engineering}, we instantiate this framework in a cavity-based micromaser, where homodyne readout gives direct access to the coherent actions. The same operational criteria and fuel classification, however, apply more broadly to collisional and repeated-interaction settings in which the working medium is another finite quantum system. The topology--thermodynamics link uncovered here is thus not specific to atom--cavity platforms: edge fuels are expected to remain \emph{heat-only} under successive coupling to a qubit working medium, whereas bulk fuels activate coherent channels.

The system under consideration in this study is illustrated in Figure~\ref{fig:figconcept}. A sequence of four-qubit atomic ensembles, prepared in a predetermined quantum state, is randomly injected into the cavity, with only one ensemble interacting with the cavity field at any given time. The arrival times of the atomic ensembles follow a Poissonian distribution. The transfer duration from the state preparation stage to the cavity is denoted as \( t_{\text{tr}} \), and once inside the cavity, each atomic ensemble interacts with the cavity field for a duration \( \tau \) before exiting.

The dynamics of the atomic ensembles interacting with the cavity field is described within the framework of the Tavis-Cummings model, using the Hamiltonian 
$H = H_\text{a} + H_\text{c} + H_{\text{int}}$
where the first two terms $H_a$ and $H_c$ represent the free Hamiltonians of the atomic ensemble and the cavity whereas the last term $H_{\text{int}}$ represent the interaction Hamiltonian~\cite{tavis1968exact}, see {\color{blue}Methods}.

Our eight-cell SSH chain is described by the Hamiltonian~\cite{su1980soliton}
\begin{equation}\label{eq:SSH}
H_{\mathrm{SSH}}=\sum_{n}\Big(t_1\,c^\dagger_{n,A}c_{n,B}+t_2\,c^\dagger_{n,B}c_{n+1,A}+\mathrm{h.c.}\Big),
\end{equation}
that supports sixteen eigenmodes  $\{\,|v_s\rangle\,\}_{s=1}^{16}$ which we use to generate micromaser fuels via $\rho_s = \frac{|v_s\rangle\!\langle v_s| } {\mathrm{Tr}(|v_s\rangle\!\langle v_s|)}$.
	To physically anchor the fuels in hardware, we design a superconducting-qubit realization that first prepares a selected sixteen-site SSH eigenstate and then transfers it to a four-qubit register using a compiled native-gate sequence (Methods). Concretely, an open SSH chain with alternating couplings $(t_1,t_2)$ is implemented on a 16-qubit array. Edge eigenstates can be prepared by adiabatic evolution within the gapped topological regime, whereas generic bulk eigenstates can be synthesized directly in the single-excitation manifold after spectroscopic identification. The subsequent transfer to the four-qubit fuel uses calibrated exchange interactions together with local single-qubit control to produce the encoded state $\sum_{k=1}^{16} c_k \ket{b(k)}$, up to known calibration phases. With native superconducting control rates compatible with present transmon hardware, the preparation-and-transfer stage remains within coherence limits, after which the same flight-and-cavity sequence used throughout is applied. This establishes that the fuels used here are not abstract encodings but arise from a concrete, platform-realizable SSH-to-fuel pipeline; alternative physical platforms for the realization of the present protocol are given in the {\color{blue}Supplementary Information Section S1}.
	
For fixed $(t_1,t_2)$ we prepare copies of a \emph{single} eigenstate $|v_s\rangle$ of $H_{\mathrm{fuel}}(t_1,t_2)$, form the fuel $\rho_s$, and inject it into the cavity in a standard repeated–interaction protocol  operated below maser threshold. 
The cavity's phase–sensitive response is described by the coherent–channel indicators $(\lambda,\xi)$, which we \emph{estimate} from homodyne/heterodyne quadratures via the steady–state (or initial–slope) relations (see {\color{blue}Supplementary Information Section S2}), yielding $\widehat{\lambda}$ and $\widehat{\xi}$. 
Sweeping the dimerization ratio $t_1/t_2$ then produces a binary, transport–free signature of edge topology (see {\color{blue}Methods}):
\begin{itemize}
	\item $\text{edge fuel:}\  |\widehat{\lambda}|, |\widehat{\xi}|\le \varepsilon \quad\text{(purely thermal Gibbs cavity)};$
	\item$\text{bulk fuel:}\  |\widehat{\lambda}|, |\widehat{\xi}|>\varepsilon \quad\text{(thermo–mechanical)},$
\end{itemize}
\noindent where $\varepsilon$ is a fixed detection threshold set by the quadrature noise floor and the independent calibrations of the proportionality constant $\mu$ (set by arrival rate and coupling/interaction time) and $\kappa$, the cavity decay rate.

The physical mechanism behind this dichotomy is already visible at the fuel level. In the topological regime, the SSH edge eigenstates are exponentially localized and inherit a highly constrained amplitude pattern from the boundary mode structure. After transfer to the four-qubit register, this structure suppresses the density-matrix elements that contribute to the one- and two-quantum coherence channels, so that the displacement and squeezing indicators vanish, $\lambda=\xi\approx 0$, while the heat-exchange sector remains active. By contrast, bulk eigenstates are spatially extended across the chain and populate the encoded register more broadly, thereby sustaining the coherences that feed both channels and yielding thermo-mechanical response.

\begin{figure}[h!]
		\includegraphics[width=0.4\linewidth]{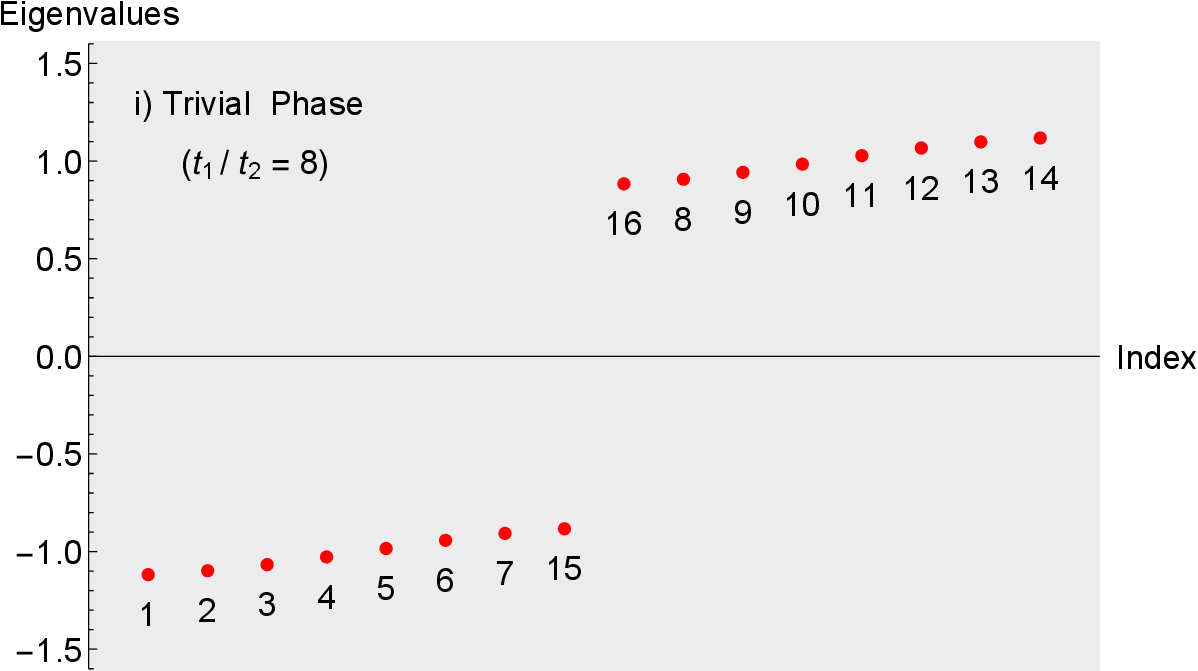}\quad 
		\includegraphics[width=0.4\linewidth]{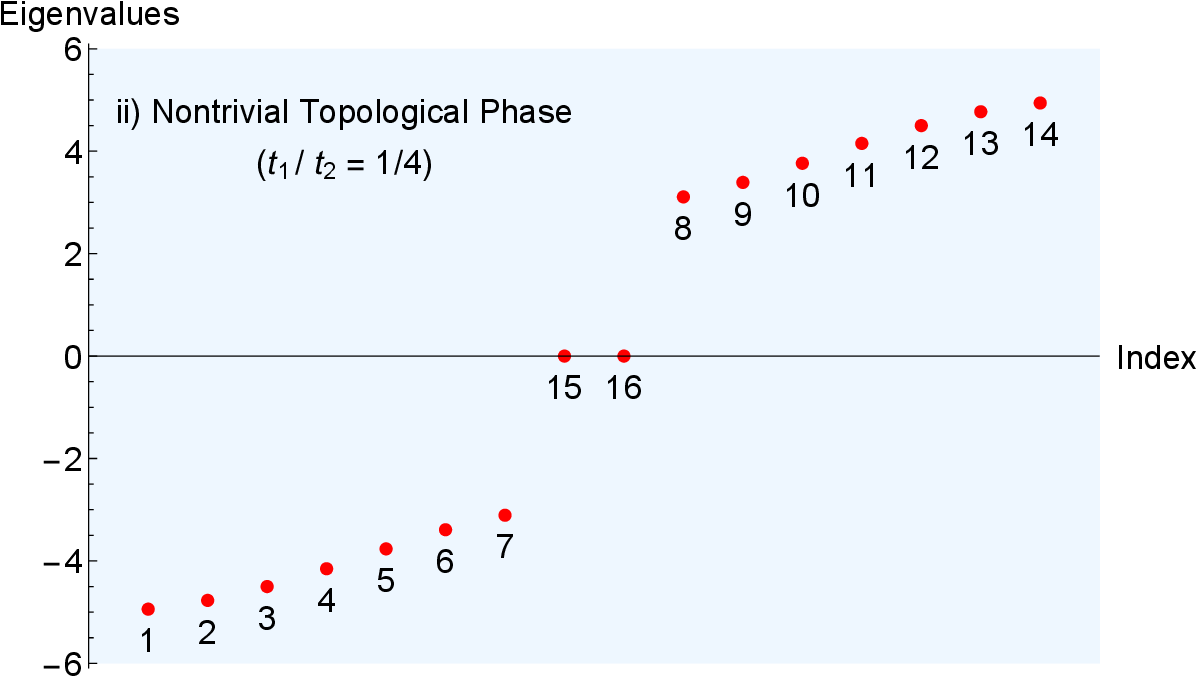} \\ \ \\
		\includegraphics[width=0.4\linewidth]{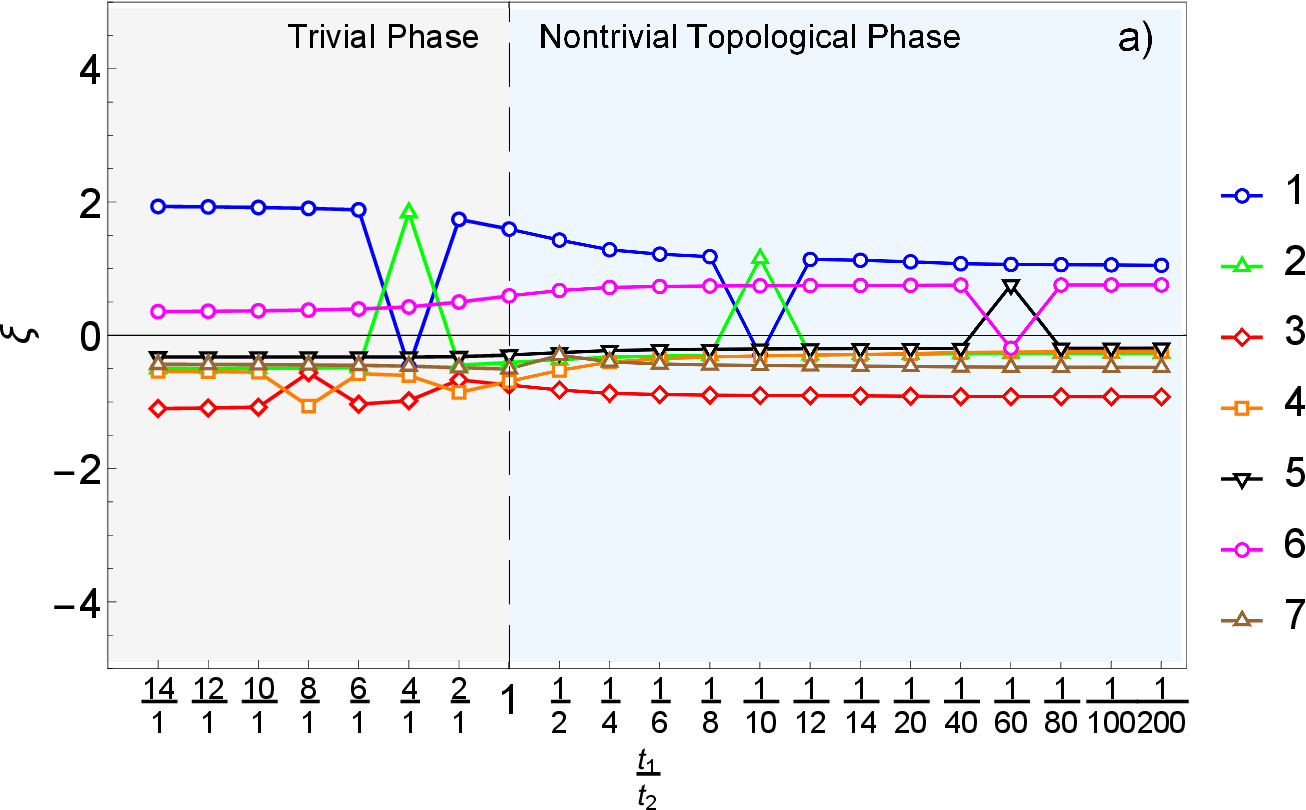}\quad
		\includegraphics[width=0.4\linewidth]{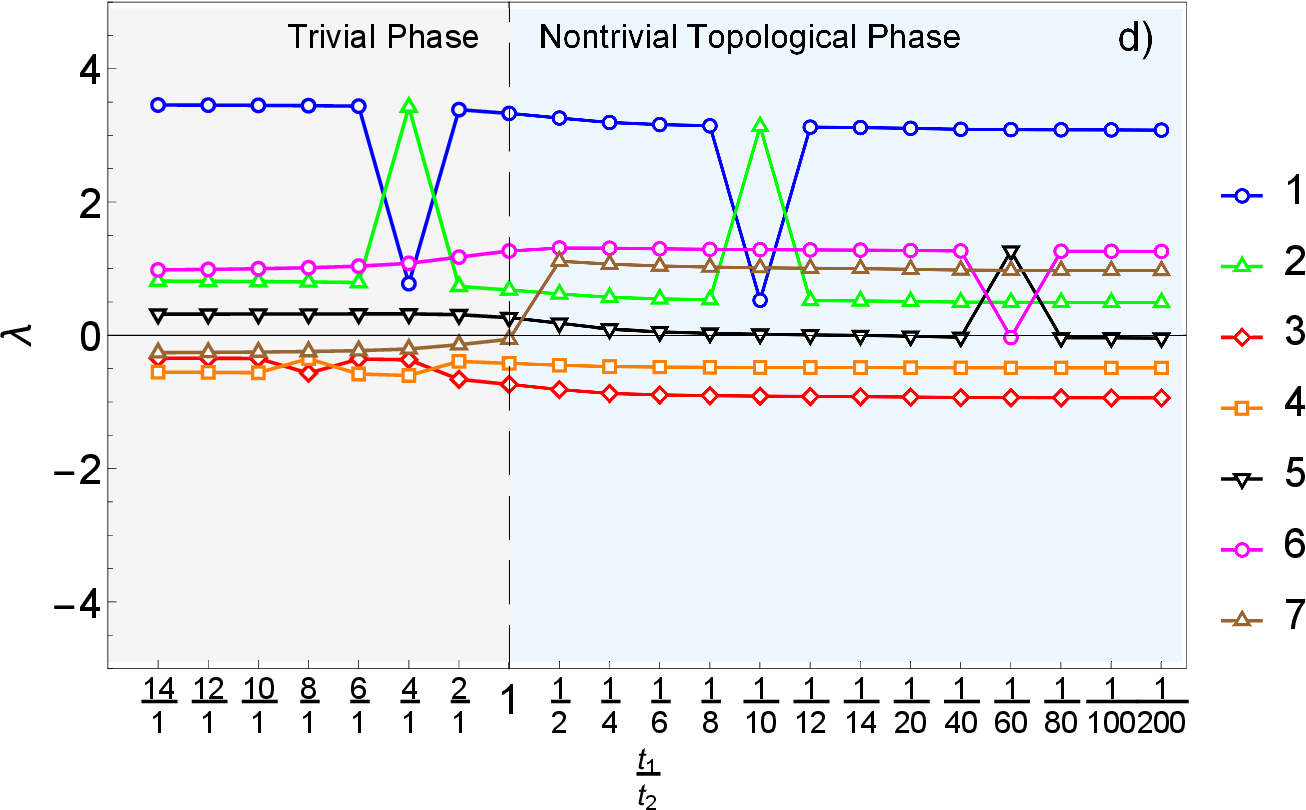} \\ \ \\
		\includegraphics[width=0.4\linewidth]{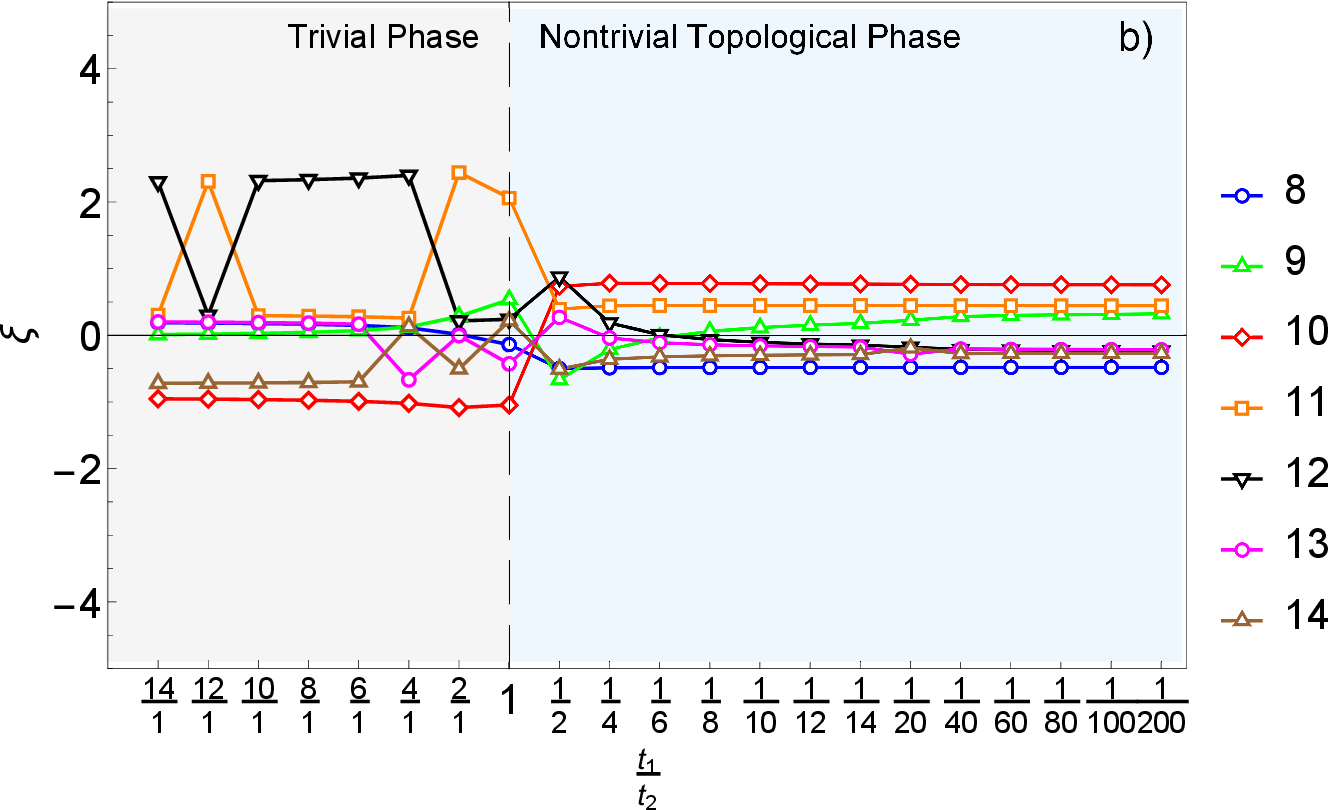}\quad
		\includegraphics[width=0.4\linewidth]{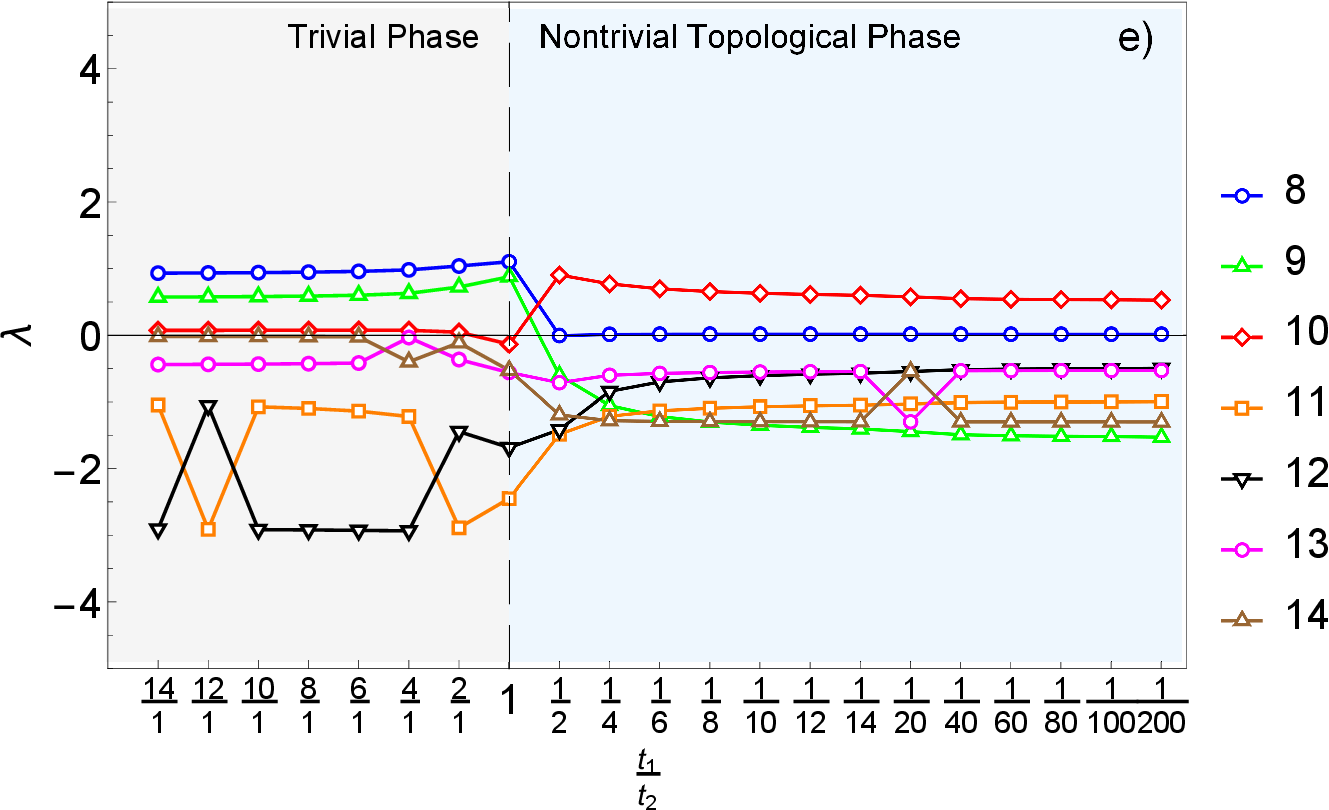} \\ \ \\
		\includegraphics[width=0.4\linewidth]{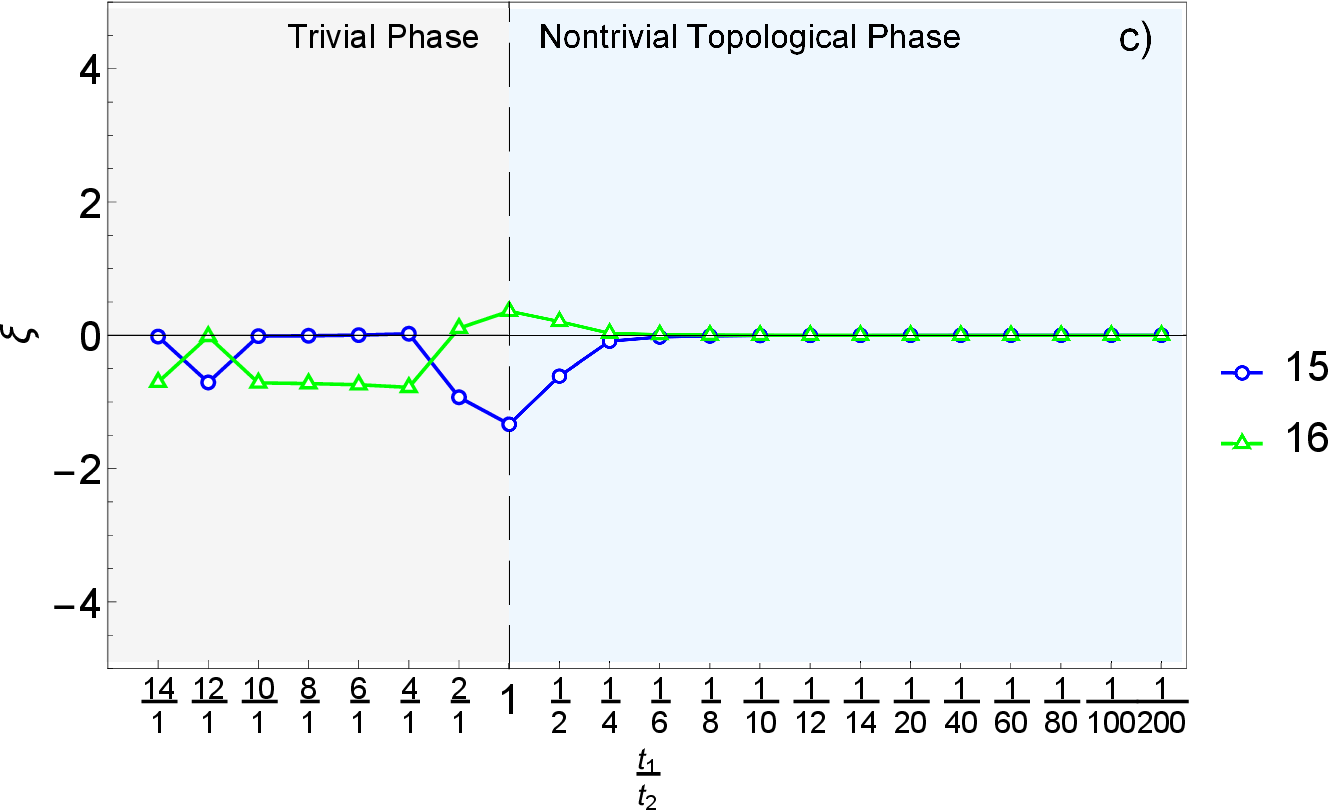}\quad
		\includegraphics[width=0.4\linewidth]{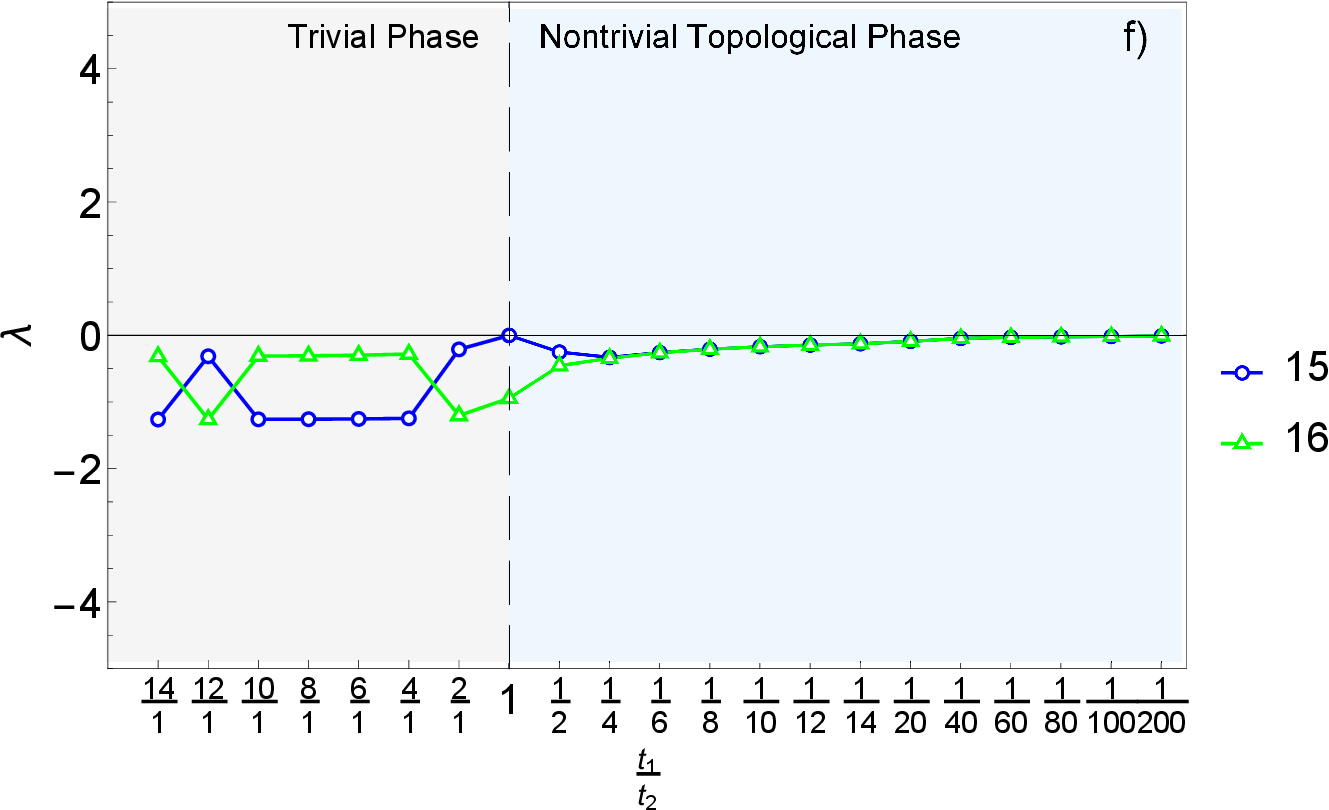} \\
		\caption{\textbf{Displacement and squeezing channels for sixteen SSH fuels.} Four-qubit encoding of an eight-cell SSH chain driving a repeated-interaction micromaser. Eigenvalue enumeration of SSH chain in i) trivial, and ii) nontrivial topological phases. For each eigenstate fuel $\rho_s = |v_s\rangle \langle v_s|$, ($s= 1,\dots,16$), data points show (a through c) the \emph{squeezing} indicator $\xi$, and (d through f) the \emph{displacement} indicator $\lambda$ as functions of the dimerization ratio $t_1/t_2$ with GADC during flight ($t_{\mathrm{tr}}=10~\mathrm{ns}$) and cavity loss included. Bulk fuels (the first 14 states) activate coherent channels  across parameters, exhibiting $\lambda\neq0$ \emph{and} $\xi\neq0$, whereas the two edge fuels (15th and 16th states) exhibit vanishing $\lambda$ and $\xi$ throughout the topological phase ($t_1<t_2$), yielding pure thermalization of the cavity.\label{fig:mainresults}}
	\end{figure}

Figure~\ref{fig:mainresults} reports, for each of the sixteen fuel states, the displacement indicator $\lambda$ and the squeezing indicator $\xi$ as functions of $t_1/t_2$ with cavity losses and generalized amplitude damping (GADC) as a relevant decoherence channel included. The screening effects remain consistent in the full micromaser: the fourteen \emph{bulk} fuels activate coherent channels across parameters (nonzero $\lambda$ and $\xi$), while the two \emph{edge} fuels in the topological phase continue to satisfy $\lambda=\xi\approx 0$ within numerical precision, thereby driving \emph{pure thermalization} of the cavity.

We emphasize that our classification method is robust to realistic cavity loss and flight-induced decoherence, relying solely on the final four-qubit fuel state. We provide a full superconducting circuit model in the {\color{blue}Supplementary Information Section S1}, detailing the preparation of the SSH eigenstate and its high-fidelity transfer to a compact register using compiled native controls built from calibrated exchange interactions and local single-qubit operations.
{\color{blue}Supplementary Information Sections S3 and S4} further show that the edge-fuel thermal response persists under depolarizing and phase-damping channels, as well as under bond disorder and moderate onsite disorder in the initial chain, respectively.

\newpage
%%%%%%%%%%%%%%%%%%%%%%%%%%%%%%%%%%%%%%%%%%%%%%%%
%\clearpage
%\begin{widetext}
	\onecolumngrid
	\onecolumngrid
%\end{widetext}
\twocolumngrid
%%%%%%%%%%%%%%%%%%%%%%%%%%%%%%%%%%%%%%%%%%%%%%%%
\ \\
\clearpage
\newpage

To test whether this behaviour is specific to topological edge structure, we also examined representative trivial lattice models across broad parameter ranges and found no analogous pure-thermalization regime within the cases explored (see {\color{blue}Supplementary Information Section S5}).

In conclusion, we have identified a direct operational bridge between topology and quantum thermodynamics: in an open eight-cell SSH chain, \emph{edge} eigenstates act as \emph{pure-thermalization fuels} for a repeated-interaction micromaser, switching off both coherent channels (displacement and squeezing), whereas all \emph{bulk} eigenstates activate them. 
The dichotomy is already visible at the fuel level via energy-basis indicators and persists under generalized amplitude damping during flight and realistic superconducting-resonator loss.

Operationally, sweeping the dimerization ratio turns the micromaser into a \emph{transport-free topology classifier}: vanishing coherent action identifies the edge regime, while simultaneous displacement and squeezing identify the bulk. 

For physical implementation of our approach, we propose a superconducting qubit route which is accessible with the current state-of-the-art superconducting technology. Beyond cavity QED, the same energy-basis criterion applies to other repeated-interaction platforms—including collisional (atom–atom / qubit–qubit) implementations—where an edge fuel remains heat-only while bulk fuels are thermo-mechanical. 

Looking forward, extending the symmetry-based conditions for coherent-channel suppression, scaling to longer chains and higher-excitation sectors, and exploring alternative topological platforms would further develop \emph{topology-assisted thermodynamic control} and broaden the utility of edge-mode fuels as passive, Gibbs-stabilizing resources.
\section{Methods}\label{sec:Methods}
In the Hamiltonian we consider, the free Hamiltonians of the atomic ensemble $H_a$ and the cavity $H_c$, and the interaction Hamiltonian $H_{\text{int}}$ are given as~\cite{tavis1968exact}
\begin{eqnarray}
	H_\text{a} &=& { \hbar \omega_a \over 2} \sum_{k=1}^4 \sigma_k^z,\\
	H_\text{c} &=& \hbar \omega_c a^{\dag} a, \\
	H_{\text{int}} &=& \hbar g \sum_{k=1}^4 ( a \sigma_k^+ + a^{\dag} \sigma_k^-),
\end{eqnarray}
\noindent where \( \sigma_k^z \), \( \sigma_k^+ \), and \( \sigma_k^- \) are the Pauli operators corresponding to the \( k^{\text{th}} \) atom, and \( a \) and \( a^{\dagger} \) denote the annihilation and creation operators of the cavity field, respectively.

We assume the atomic transition frequency \( \omega_a \) is in resonance with the cavity frequency \( \omega_c \). In the interaction picture, the combined system of the atomic ensemble and the cavity undergoes unitary evolution according to \( U(\tau) = \text{exp}(-i H_{\text{int}} \tau) \). Given that the interaction involves repeated encounters between atomic ensembles and the cavity field, we focus on the cavity state after the passage of the \( j^{\text{th}} \) atomic ensemble, which enters the cavity at \( t_j \) and exits at \( t_j + \tau \). The cavity's state \( \rho_c \) is obtained by tracing out the atomic degrees of freedom
\begin{equation}
	\rho_c(t_j + \tau) = \text{tr}_a [ U(\tau) \rho_a \otimes \rho (t_j) U^{\dagger}(\tau) ],
	\label{eq:TCMthermalOperation}
\end{equation}
\noindent and governed by the master equation~\cite{filipowicz1986theory,liao2010single}
\begin{equation}\label{eq:mastereq1}
	\dot{\rho}_c = p\!\left[\sum_{i,j=1}^{16} a_{ij}\sum_{n=1}^{16} U_{ni}(\tau)\,\rho_c\,U^\dagger_{nj}(\tau)-\rho_c\right],	
\end{equation}
where the elements \(a_{ij}\) represent the density matrix components of the atomic ensemble. This master equation can also be expressed in the operational form
\begin{equation}\label{eq:mastereq2}
	\dot{\rho}_c \approx -i[H_{\mathrm{eff}},\rho_c]+\mathbb{L}_s[\rho_c]+\mathbb{L}[\rho_c],
\end{equation}
\noindent which is valid under the standard weak-coupling/short-interaction assumptions: $\mu\tau\!\ll\!1$ and dilute arrivals ($p\tau\!\ll\!1$), Born–Markov + secular/RWA for both ancilla–cavity and cavity–bath couplings, and fresh, uncorrelated ancillas between shots.

The coherent drive which is a common displacement is described by the effective Hamiltonian
\begin{equation}\label{eq:CoherentDrive}
	H_{\text{eff}} = pg\tau (\lambda a^{\dagger} + \lambda^{*} a),
\end{equation}
and the squeezing process is described by the Lindbladian
\begin{equation}
	\mathbb{L}_s \rho = \mu (\xi \mathbb{L}^e_s + \xi^{*} \mathbb{L}^d_s )
\end{equation} 
with the excitation and de-excitation squeezing Lindbladians $\mathbb{L}^e_s$ and $\mathbb{L}^d_s$, respectively. 

The heat exchange process is described by
\begin{equation}
	\mathbb{L}\rho = \mu (\frac{r_e}{2}\mathbb{L}_e + \frac{r_g}{2}\mathbb{L}_d ) 
\end{equation}
with incoherent excitation and de-excitation Lindbladians $\mathbb{L}_e$ and $\mathbb{L}_d$, respectively. 
\begin{figure}[t!]
	\includegraphics[width=0.55\linewidth]{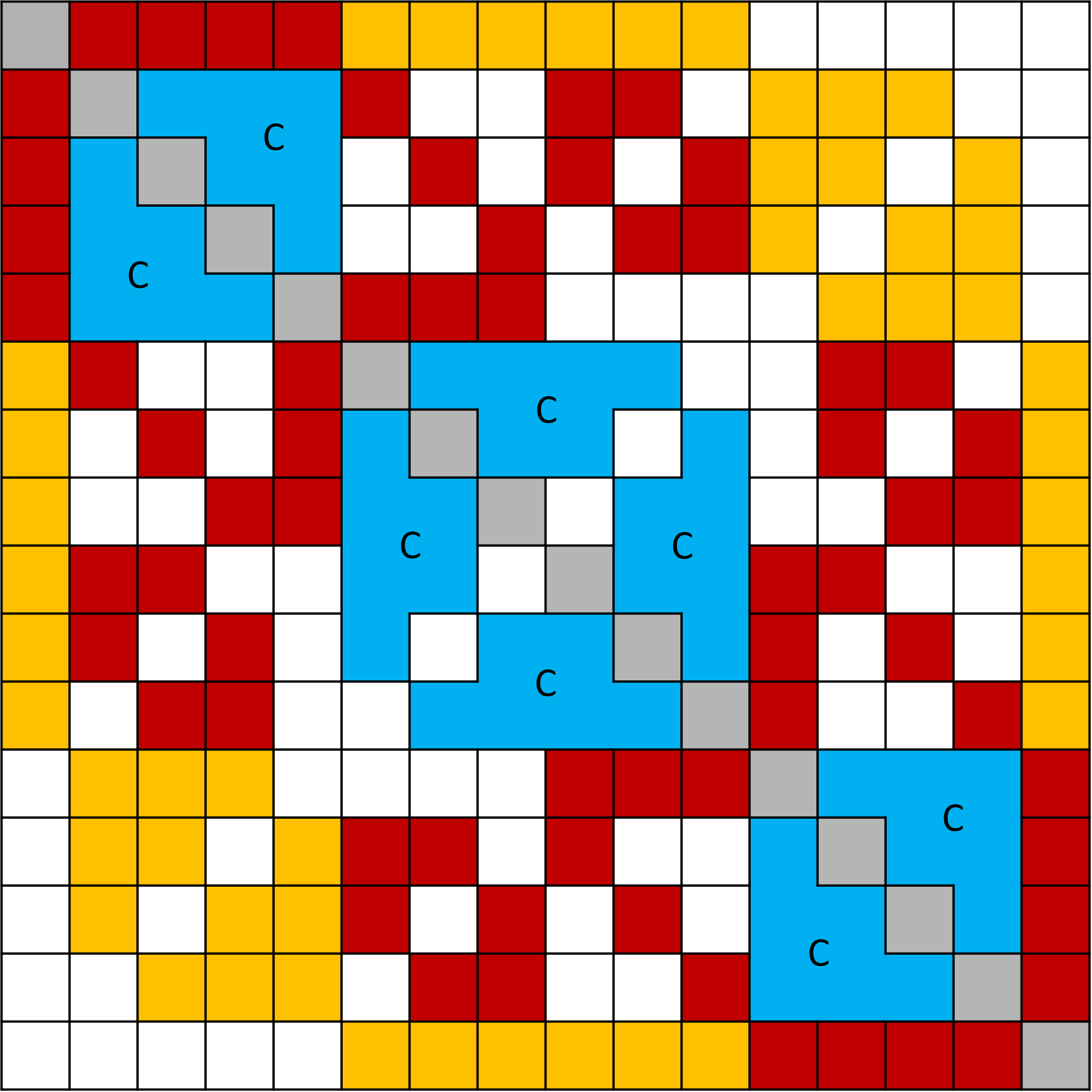}
	\caption{\textbf{Illustration of the density matrix $\rho$ of the four-qubit fuel system in energy basis.} Blue blocks represent the coherence terms driving the incoherent heat exchange process, and red and yellow terms represent the terms contributing to coherence injection, (summing up to form $\lambda$) and squeezing (summing up to obtain $\xi$), respectively. Gray diagonal and white terms are the populations and ineffective coherences, respectively~\cite{manatuly2019collectively}. $\lambda=\xi=0$ satisfy the pure thermalization condition.\label{fig:figMatrix}}
\end{figure} 
Pure thermalization means the dynamics reduces to the thermal Lindbladian, $\dot{\rho}_c\approx\mathbb{L}\rho_c$. 
In the moment equations, the fuel contributes source terms $\propto\mu\lambda$ to $\dot{\langle a\rangle}$ and $\propto\mu\xi$ to $\dot{\langle a^2\rangle}$; any nonzero $\lambda$ or $\xi$ drives phase-sensitive steady values and thus precludes a Gibbs fixed point. 
Barring nongeneric fine-tuned cancellations, one therefore has the operational equivalence
$\lambda=\xi=0$ for pure thermalization.

In the energy eigenbasis of the four–qubit fuel, the cavity response separates into three disjoint groups of matrix elements as illustrated in Figure~\ref{fig:figMatrix}: (i) the \emph{displacement} (single–quantum) coherences, summed by \(\lambda\); (ii) the \emph{squeezing} (two–quantum) coherences, summed by \(\xi\); and (iii) the \emph{heat–exchange} coherences $C$ (together with the populations), which set the Gibbs temperature. 
Terms forming $\lambda$ and $\xi$ are provided in the {\color{blue}Supplementary Information Section S6}.

\textbf{Steady-state characterization.}
Below the maser threshold ($r_e<r_g$), the evolution with $\lambda=\xi=0$ reduces to a
thermalizing Liouvillian and yields a Gibbs steady state with temperature set by
the competition of heat-exchange rates and the thermal bath,
\begin{equation}
	\frac{r_e+\bar{n}_{\mathrm{th}}\kappa/\mu}{r_g+(\bar{n}_{\mathrm{th}}+1)\kappa/\mu}
	= e^{-\hbar\omega_c/k_B T_{\mathrm{cav}}}\!,
\end{equation}
equivalently
\begin{equation}
	T_{\mathrm{cav}}=\frac{\hbar\omega_c}{k_B}\!\left[\ln\!\left(\frac{R+\delta+2C+2(\bar{n}_{\mathrm{th}}+1)\kappa/\mu}
	{R-\delta+2C+2\bar{n}_{\mathrm{th}}\kappa/\mu}\right)\right]^{-1}\!,
\end{equation}
with $R=R_g+R_e=4$, $\delta=R_g-R_e$, and $C$ the aggregate heat-exchange coherence
(see \cite{dag2019temperature,tuncer2019work} for definitions).
Evaluations using the GADC-modified fuels confirm that \emph{edge} fuels remain in the
purely thermal sector (vanishing coherent channels) while \emph{bulk} fuels yield
thermo-mechanical response.

\textbf{Realistic micromaser dynamics.}
We propagate the cavity under repeated interactions with Poissonian injections,
including (i) generalized amplitude damping (GADC) during a finite flight time
$t_{\mathrm{tr}}$ at ambient temperature $T_{\mathrm{env}}$, with strength
$q=1-\exp[-\gamma t_{\mathrm{tr}}(1+2\bar{n}_{\mathrm{th}})/2]$ and thermal weights
$\alpha=(\bar{n}_{\mathrm{th}}+1)/(2\bar{n}_{\mathrm{th}}+1)$,
$\beta=\bar{n}_{\mathrm{th}}/(2\bar{n}_{\mathrm{th}}+1)$ for amplitude damping (ADC) and amplitude amplifying  (AAC) channels, respectively, and (ii) cavity damping
$\mathbb{L}_c[\rho_c]=\tfrac{\kappa}{2}\!\left((\bar{n}_{\mathrm{th}}+1)\mathbb{L}_d[\rho_c]
+\bar{n}_{\mathrm{th}}\mathbb{L}_e[\rho_c]\right)$, with
$\bar{n}_{\mathrm{th}}=(e^{\hbar\omega_c/k_B T_{\mathrm{env}}}-1)^{-1}$.
Kraus operators for the GADC, which is a combination of ADC and AAC are given as 
$\{\sqrt{\alpha} \left(\Ket{0}\Bra{0} + \sqrt{1-q} \Ket{1}\Bra{1} \right)$, 
$\sqrt{\beta} \left(\sqrt{1-q} \Ket{0}\Bra{0} + \Ket{1}\Bra{1} \right)$, 
$\sqrt{\alpha q} \Ket{0}\Bra{1}$, 
$\sqrt{\beta q} \Ket{1}\Bra{0} \}$. We adopt superconducting-resonator parameters consistent with
Refs.~\cite{dag2019temperature,tuncer2019work}:
$\omega_c/2\pi=10~\mathrm{GHz}$, $T_{\mathrm{env}}\!\approx\!160~\mathrm{mK}$
($\bar{n}_{\mathrm{th}}\!\approx\!0.05$), atomic relaxation $\gamma/2\pi=1~\mathrm{MHz}$,
and a representative micromaser scaling $\kappa/\mu=1$. For the noise stress-test we use a realistic flight time, $t_{\mathrm{tr}}=10~\mathrm{ns}$. 

\section*{Acknowledgements} F.O. acknowledges financial support from Tokyo International University Personal Research Fund and Special Grant-in-Aid for Research Work. S.K.O and A.L. acknowledge support from Air Force Office of Scientific Research (AFOSR) Multidisciplinary University Research Initiative (MURI) award no. FA9550-21-1-0202.

%\defaultbibliography{C:/Users/seval/Dropbox/OQuL/OQuLBib/OQuL} % Your main .bib file
\bibliography{topo4thermo-20260717-arXiv}
\clearpage % Start supplementary material on a new page

%\label{sec:supplementary}
% Define commands to prefix supplementary content
\makeatletter
\renewcommand{\thesection}{S\@arabic\c@section}
\renewcommand{\thesubsection}{S\thesection.\@arabic\c@subsection}
\renewcommand{\thefigure}{S\@arabic\c@figure}
\renewcommand{\thetable}{S\@arabic\c@table}
\renewcommand{\theequation}{S\@arabic\c@equation}
\renewcommand{\bibnumfmt}[1]{[S#1]}
\renewcommand{\citenumfont}[1]{S#1}
\makeatother

\clearpage
\newpage
\onecolumngrid
% Set up a new bibliography unit for supplementary material
\begin{bibunit}[apsrev4-2] % Use the same or a different style if needed

\section{Supplementary Information}\label{sec:Supplementary}

\section*{S1. SUPERCONDUCTING-QUBIT REALIZATION: SSH PREPARATION AND COMPILED TRANSFER TO A QUBIT FUEL}\label{sec:S1}

\subsection{S1.a. Device model and native controls}
We consider fixed-frequency or weakly tunable transmons $Q_i$ coupled by either a bus resonator or dedicated tunable couplers. In the dispersive regime the relevant few-level Hamiltonian reduces, after a Schrieffer–Wolff elimination, to an effective qubit model with always-on $ZZ$ and parametric/exchange interactions~\cite{blais2021circuit,kjaergaard2020superconducting}:
\begin{equation}
	H_{\mathrm{ctrl}}(t)
	=\sum_i \frac{\omega_i(t)}{2} Z_i
	+\sum_{i<j}\!\Big[ J_{ij}(t)\,\frac{X_iX_j+Y_iY_j}{2}
	+\zeta_{ij}(t)\,Z_i Z_j\Big]
	+\sum_i \big(\Omega_i^x(t) X_i+\Omega_i^y(t) Y_i\big).
	\label{eq:ctrl}
\end{equation}
Parametric flux modulation of a tunable coupler at the frequency detuning between qubits $i,j$ makes $J_{ij}(t)$ time-programmable and realizes high-fidelity iSWAP/partial-iSWAP gates \cite{caldwell2018parametrically,mundada2019suppression,sung2021realization}. Typical values are $J_{ij}/2\pi\simeq 5$–$20$ MHz; single-qubit Rabi rates $\Omega/2\pi\simeq 20$–$50$ MHz.

\subsection*{S1.b. A physical SSH chain on superconducting hardware}
We consider implementing an \emph{actual} SSH Hamiltonian on a linear array of 16 transmons (or 16 microwave resonators with embedded qubits), with alternating couplings $(t_1,t_2)$:
\begin{equation}
	H_{\mathrm{SSH}}(t_1,t_2)
	=\sum_{n=1}^{8}\Big[t_1\,\sigma^+_{A_n}\sigma^-_{B_n}
	+t_2\,\sigma^+_{B_n}\sigma^-_{A_{n+1}}+\mathrm{h.c.}\Big]
	+\sum_{i}\frac{\tilde{\omega}_i}{2} Z_i,
	\label{eq:Hssh16}
\end{equation}
where $(A_n,B_n)$ denote the two sites in cell $n$, and open boundaries set $A_{9}\equiv\varnothing$. In the single-excitation sector spanned by $\{|k\rangle\}_{k=1}^{16}$ (one excitation localized on site $k$), \eqref{eq:Hssh16} is a $16\times16$ tight-binding matrix with alternating off-diagonal elements $t_1,t_2$ and two mid-gap edge eigenstates for $t_1<t_2$.

	\textit{Preparation of target eigenstates on the 16-site chain.} The preparation route depends on the chosen eigenmode. For an edge mode, one may initialize a boundary-localized single excitation and adiabatically tune the couplings within the gapped topological regime while avoiding the critical point $t_1=t_2$. Concretely, we use
	\begin{equation}
		t_1(s)=t \big[1-\delta(s)\big],\qquad
		t_2(s)=t \big[1+\delta(s)\big],\qquad s\in[0,1],
	\end{equation}
	with $\delta(0)=\delta_i>0$, $\delta(1)=\delta_f>0$, and $\dot{\delta}/\Delta^2\ll 1$, where $\Delta(s)$ is the instantaneous single-excitation gap. This provides a high-fidelity route to the desired edge eigenstate. For a generic bulk mode, we do not rely on this adiabatic path; instead, after spectroscopic identification of the target eigenvector of $H_{\mathrm{SSH}}(t_1,t_2)$, we synthesize the corresponding state $\ket{v_s}=\sum_k c_k\ket{k}$ directly in the single-excitation manifold using the native controls of Eq.~(\ref{eq:ctrl}), namely calibrated exchange couplings together with local phases and rotations. Thus the adiabatic route is intended for edge-state preparation, whereas generic bulk-state preparation uses standard compiled state synthesis on the 16-qubit array. Because Eq.~(S13) provides site-resolved phase control together with programmable exchange couplings on a connected array, standard compiled single-excitation state-synthesis methods allow preparation of an arbitrary target eigenvector in the one-excitation manifold. In both cases the prepared state can be verified by single-excitation spectroscopy or tomography before transfer to the four-qubit fuel register.

\subsection*{S1.c. Compiled native-gate transfer from the 16-site SSH chain to a 4-qubit fuel register}

Let a separate four-qubit register be initialized in the state $|0000\rangle_R$. For a selected SSH eigenstate prepared on the 16-site chain,
\begin{equation}
	|v_s\rangle_{\mathrm{chain}}
	=\sum_{k=1}^{16} c_k |k\rangle_{\mathrm{chain}} .
	\tag{S16}
\end{equation}
We compile a state-specific unitary $U_s$ acting on the joint Hilbert space of the chain and the register such that
\begin{equation}
	U_s \left(
	|v_s\rangle_{\mathrm{chain}} \otimes |0000\rangle_R
	\right)
	=|0\rangle_{\mathrm{chain}} \otimes |\widetilde v_s\rangle_R .
	\tag{S17}
\end{equation}
Here $|0\rangle_{\mathrm{chain}}$ denotes the zero-excitation state of the 16-site chain. The encoded four-qubit state is
\begin{equation}
	|\widetilde v_s\rangle_R
	=\sum_{k=1}^{16} c_k |b(k)\rangle_R .
	\tag{S18}
\end{equation}
Here $b(k)$ denotes a fixed four-bit label assigned bijectively to the chain-site index $k$, so that the 16 sites are mapped one-to-one onto the 16 computational-basis states of the four-qubit register.
Thus, the final register state reproduces the amplitudes and phases of the selected SSH eigenstate in the computational basis of the register, up to known calibration phases that can be corrected virtually.

The transfer is implemented using the native controls introduced in Eq.(\ref{eq:ctrl}). Exchange interactions are used to move amplitude coherently between selected chain sites and register qubits, while local single-qubit phases and rotations, together with native two-qubit gates when needed, complete the encoding on the register. Partial-iSWAP operations therefore serve as routing primitives within the compiled circuit, but they do not by themselves realize the full map $|k\rangle_{\mathrm{chain}}\mapsto|b(k)\rangle_R$, since exchange-only dynamics preserve the total excitation number.

Accordingly, the protocol used here is state-specific rather than state-agnostic. For each selected eigenstate $|v_s\rangle_{\mathrm{chain}}$, the pulse sequence is compiled using the coefficients ${c_k}$ obtained from spectroscopy and numerical diagonalization of the implemented SSH Hamiltonian. Once compiled, however, the sequence deterministically prepares the same encoded four-qubit fuel state $|\widetilde v_s\rangle_R$.

A convenient implementation is to first transfer amplitude from the SSH chain to the register using calibrated exchange pulses, and then complete the final encoding by local register operations. The precise gate decomposition is not unique and depends on the hardware connectivity and compilation strategy. For the purposes of the present work, this is sufficient, because the micromaser analysis depends only on the final four-qubit fuel state and not on a unique circuit synthesis.

After the transfer, the chain is decoupled and the register state $|\widetilde v_s\rangle_R$ is used as the fuel in the repeated-interaction protocol described in the main text. In this way, the superconducting implementation provides a concrete physical route from a prepared SSH eigenstate on the 16-site chain to the four-qubit fuel register employed in our micromaser analysis.

	\subsection*{S1.d. Timing and error budget (typical superconducting parameters)}
	For exchange rates $g/2\pi = 10$--$20$ MHz, partial-iSWAP primitives with angle $\theta$ take $\tau = \theta/g \simeq 6$--$25$ ns, while single-qubit phase and rotation operations are faster. Because the state-transfer protocol of Sec.~S1.c is compiled and state-specific, the total preparation-and-transfer time depends on the chosen target eigenstate, hardware connectivity, and compilation strategy. For representative superconducting control parameters, however, the overall transfer remains well within typical transmon coherence times. With transmon $T_1,T_2^\ast \sim 50$--$150~\mu\mathrm{s}$ and calibrated echo sequences to suppress residual $ZZ$ couplings, the dominant imperfections are expected to arise from control errors and dephasing at the $10^{-3}$--$10^{-2}$ level, which is compatible with the requirements of the present proposal~\cite{blais2021circuit,kjaergaard2020superconducting}.

\subsection*{S1.e. From physical SSH to micromaser fuel}
After compression, we decouple the chain (turn couplers off), optionally verify the 4-qubit state by single-qubit tomography, then launch the \emph{fuel} (the 4-qubit register) into the repeated-interaction micromaser protocol as specified in the main text. Because the compiled transfer prepares the encoded register state $|\widetilde v_s\rangle_R$ corresponding to the selected SSH eigenstate, up to known calibration phases that are corrected during compilation, the fuel density matrix used in the micromaser protocol is $\rho_s = |\widetilde v_s\rangle_R\langle \widetilde v_s|$.

\subsection*{S1.f. Remarks and generality}
This superconducting implementation is one concrete route. The same ``prepare on a \emph{native} SSH device, then \emph{transfer to a compact register using native controls}'' blueprint carries over to trapped-ion spin chains (Mølmer–Sørensen exchange)~\cite{sorensen1999quantum,leibfried2003quantum,leibfried2003experimental} and to ultracold-atom optical superlattices, where alternating tunnelings are realized in double-well lattices~\cite{sebby2006lattice,lee2007sublattice,anderlini2007controlled,gross2017quantum} and phase-programmable Raman/laser-assisted tunneling enables controlled amplitude routing~\cite{miyake2013realizing,aidelsburger2013realization,sias2008observation}. Interfacing the lattice to a site-addressable \emph{register} is natural using optical tweezer arrays~\cite{barredo2016atom,kaufman2021quantum}. Our micromaser analysis depends only on the final 4-qubit fuel; thus any platform that realizes (i) an SSH eigenstate in the single-excitation sector and (ii) a controllable unitary transfer to a 4-qubit register suffices.

\section{S2. Extraction of coherence injection ($\lambda$) and squeezing ($\xi$) from cavity quadratures}\label{sec:S2}
We read out the cavity by phase-sensitive homodyne/heterodyne detection of the field quadratures $X=(a+a^\dagger)/2$ and $P=(a-a^\dagger)/(2i)$~\cite{walls2025quantum, raimond2006exploring, gardiner2004quantum}. 
Below maser threshold and in the weak-interaction regime used here, linearized micromaser equations give
\begin{equation}
	\frac{d}{dt}\langle a\rangle = -\frac{\kappa}{2}\langle a\rangle \;+\; \mu\,\lambda\,,\qquad
	\frac{d}{dt}\langle a^2\rangle = -\kappa\,\langle a^2\rangle \;+\; 2\mu\,\xi\,,
\end{equation}
(see, e.g., micromaser treatments in \cite{filipowicz1986theory,carmichael2013statistical,gardiner2004quantum}). In steady state (or by exponential fits of the transient) this yields
\begin{equation}
	\alpha_{\mathrm{ss}}\equiv \langle a\rangle_{\mathrm{ss}}=\frac{2\mu}{\kappa}\,\lambda \;+\; \mathcal{O}(\mu^2),\qquad
	m_{\mathrm{ss}}\equiv \big\langle a^2\big\rangle_{\mathrm{ss}}-\langle a\rangle_{\mathrm{ss}}^2=\frac{2\mu}{\kappa}\,\xi \;+\; \mathcal{O}(\mu^2).
\end{equation}
Operationally, we estimate $\lambda$ from the coherent amplitude $\alpha_{\mathrm{ss}}$ (mean of the complex IQ record) via $\widehat{\lambda}=(\kappa/2\mu)\,\alpha_{\mathrm{ss}}$, and $\xi$ from the \emph{anomalous} second moment $m_{\mathrm{ss}}$, which can be reconstructed from the quadrature covariance matrix $V$,
\begin{equation}
	m_{\mathrm{ss}}=\tfrac{1}{2}\big(\langle X^2\rangle-\langle P^2\rangle\big)\;-\;\tfrac{i}{2}\,\langle XP{+}PX\rangle,
\end{equation}
using standard Gaussian-state relations \cite{leonhardt1997measuring, weedbrook2012gaussian}. Thermal backgrounds are calibrated by measuring $V$ for a reference thermal state at the same bath temperature ($\bar n_{\mathrm{th}}$); this sets $\langle a\rangle{=}0$ and $m{=}0$ for the ``heat-only'' baseline~\cite{walls2025quantum, raimond2006exploring}. The proportionality constants $\mu$ (set by arrival rate and coupling/interaction time) and $\kappa$ are taken from independent calibrations (e.g., resonator ringdown / linewidth and injection-rate sweeps) \cite{blais2021circuit,raimond2006exploring}. As a short-time alternative that avoids steady-state fitting, one can use initial slopes,
$({d}\langle a\rangle/{dt})|_{t\to 0^+}=\mu\lambda$ and $({d}\langle a^2\rangle/{dt})|_{t\to 0^+}=2\mu\xi$ \cite{carmichael2013statistical, gardiner2004quantum}.

%\newpage
\ \\ \ \\

\section{S3. Robustness of the edge-fuel response under decoherence}\label{sec:S3}

In this section we test whether the edge-fuel pure-thermalization signature survives decoherence channels beyond GADC by applying standard single-qubit noise maps independently to each qubit of the four-qubit fuel.

\newpage
\subsection*{S3.a. Phase-damping channel}

\begin{figure}[b!]
	\includegraphics[width=0.45\linewidth]{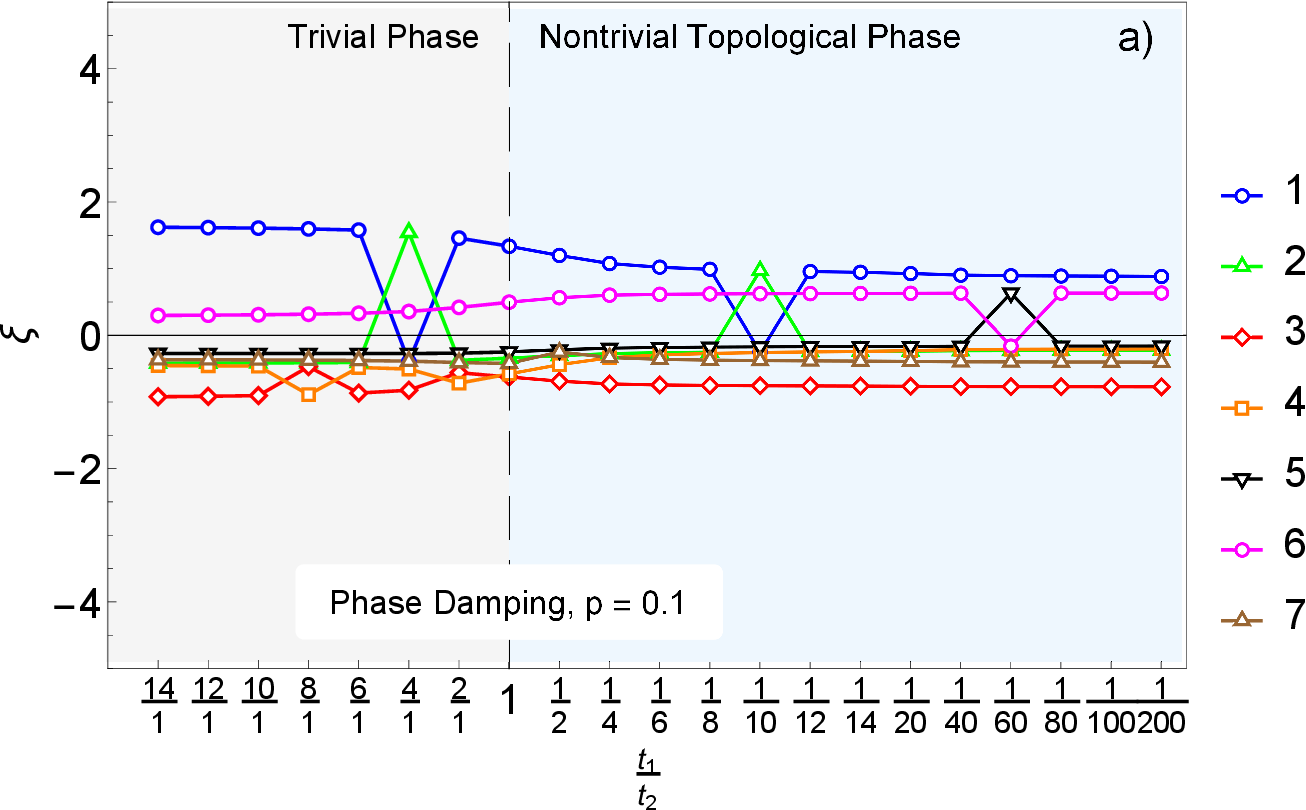}\quad
	\includegraphics[width=0.45\linewidth]{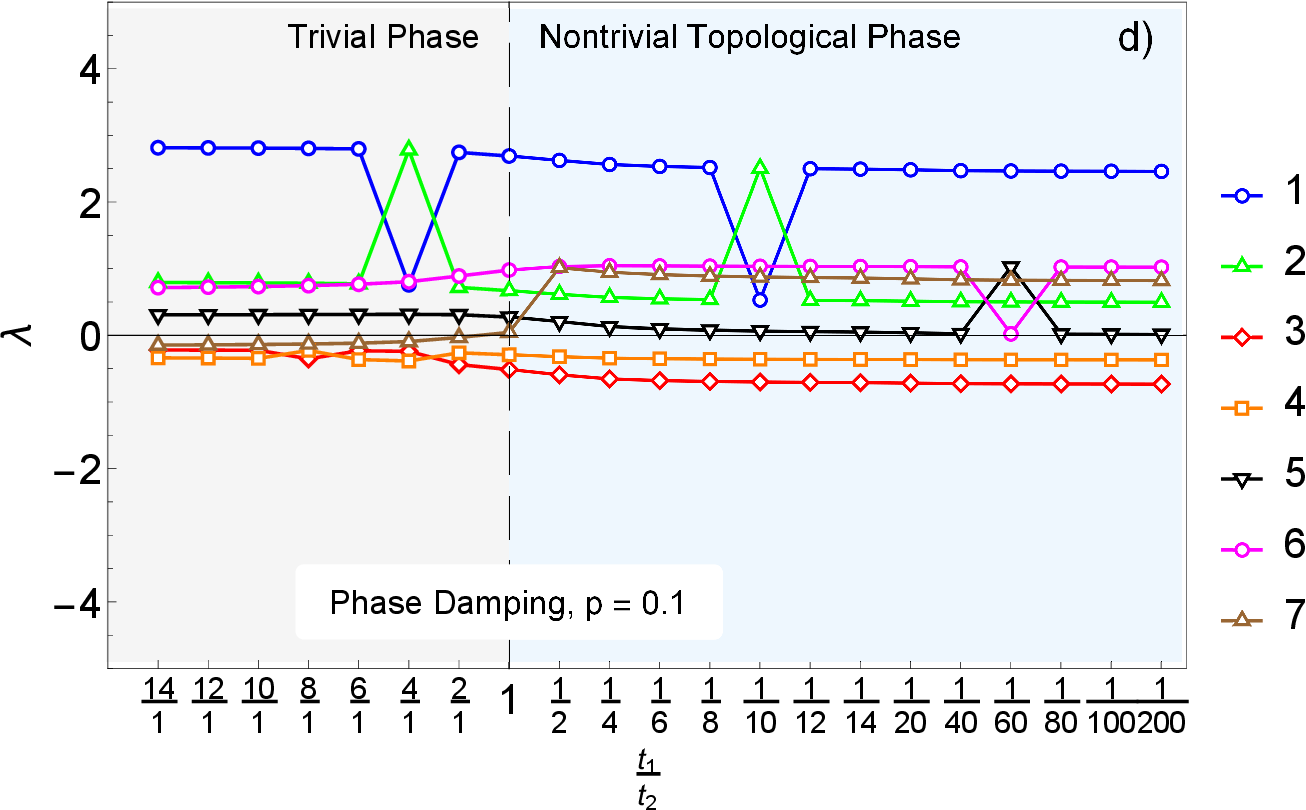} \\ \ \\
	\includegraphics[width=0.45\linewidth]{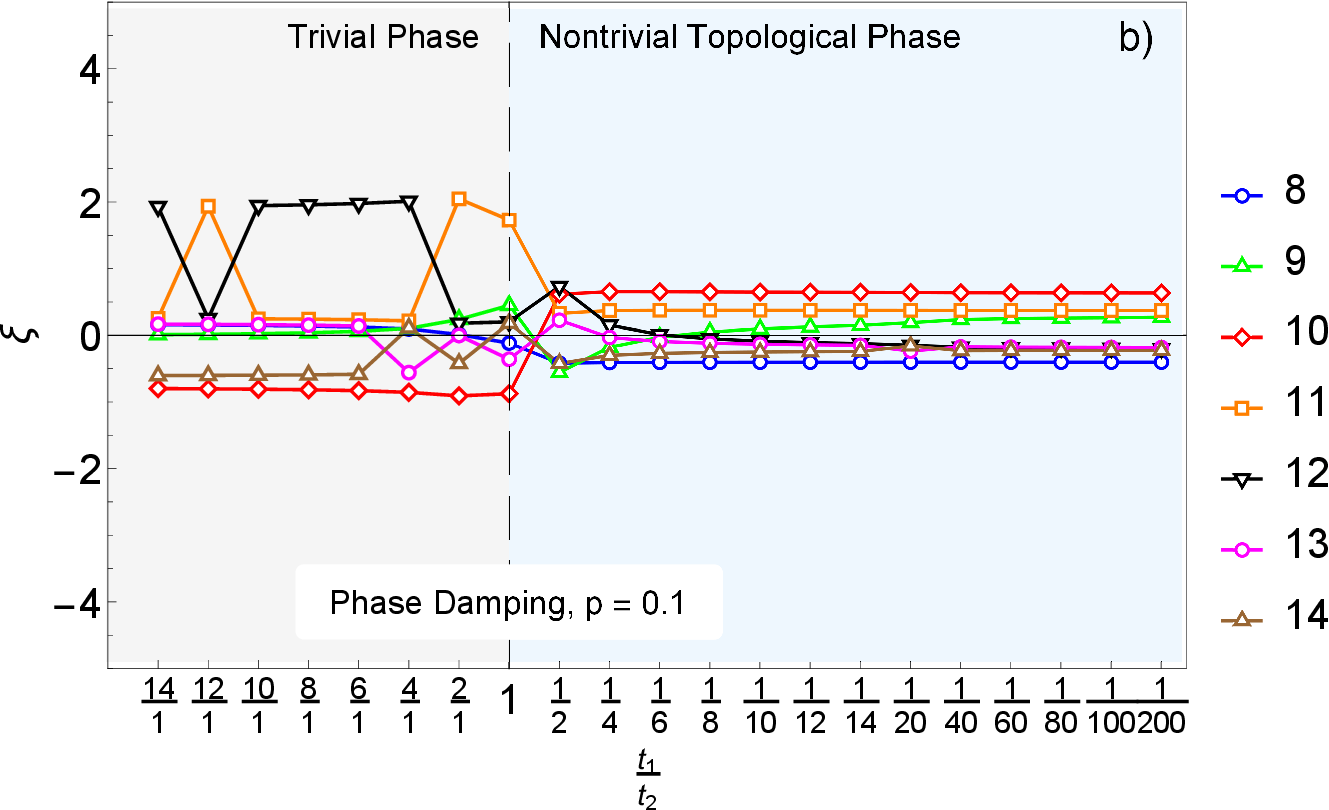}\quad
	\includegraphics[width=0.45\linewidth]{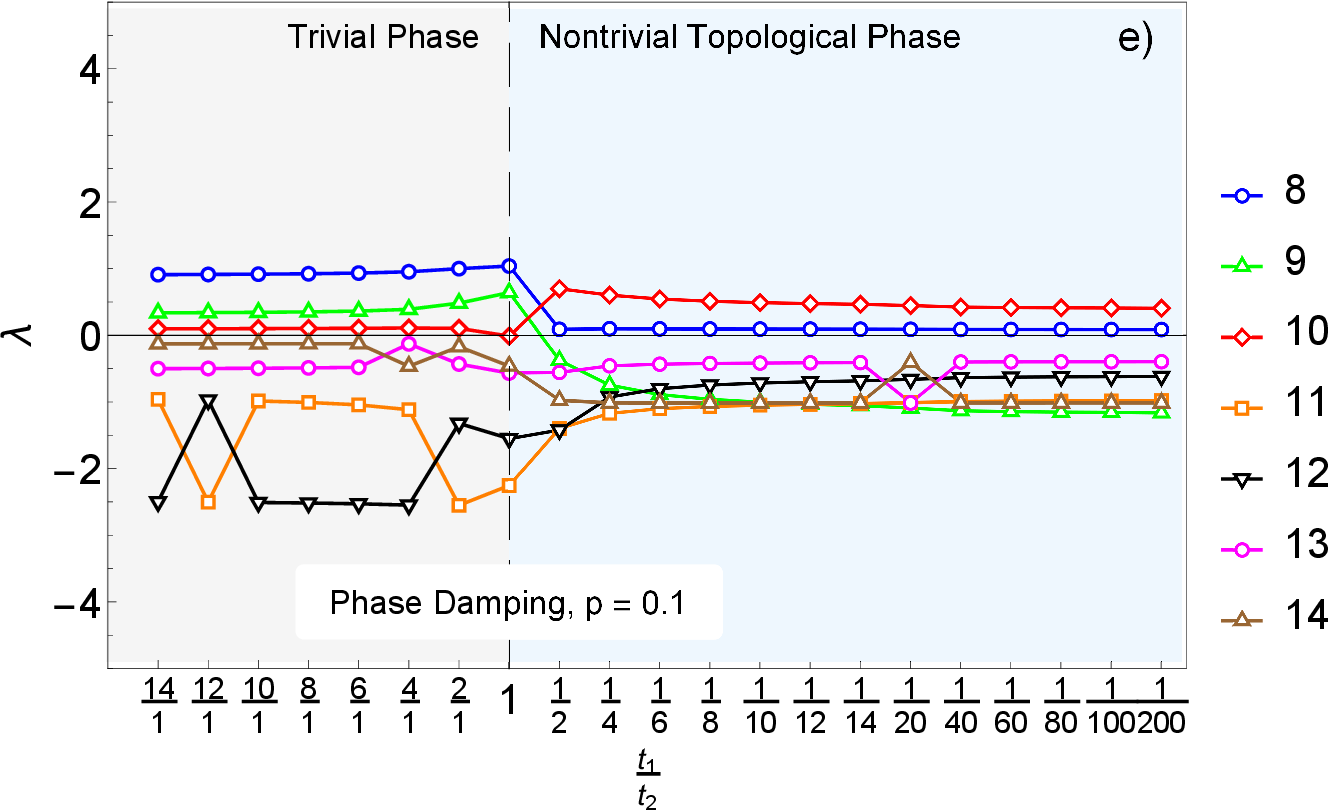} \\ \ \\
	\includegraphics[width=0.45\linewidth]{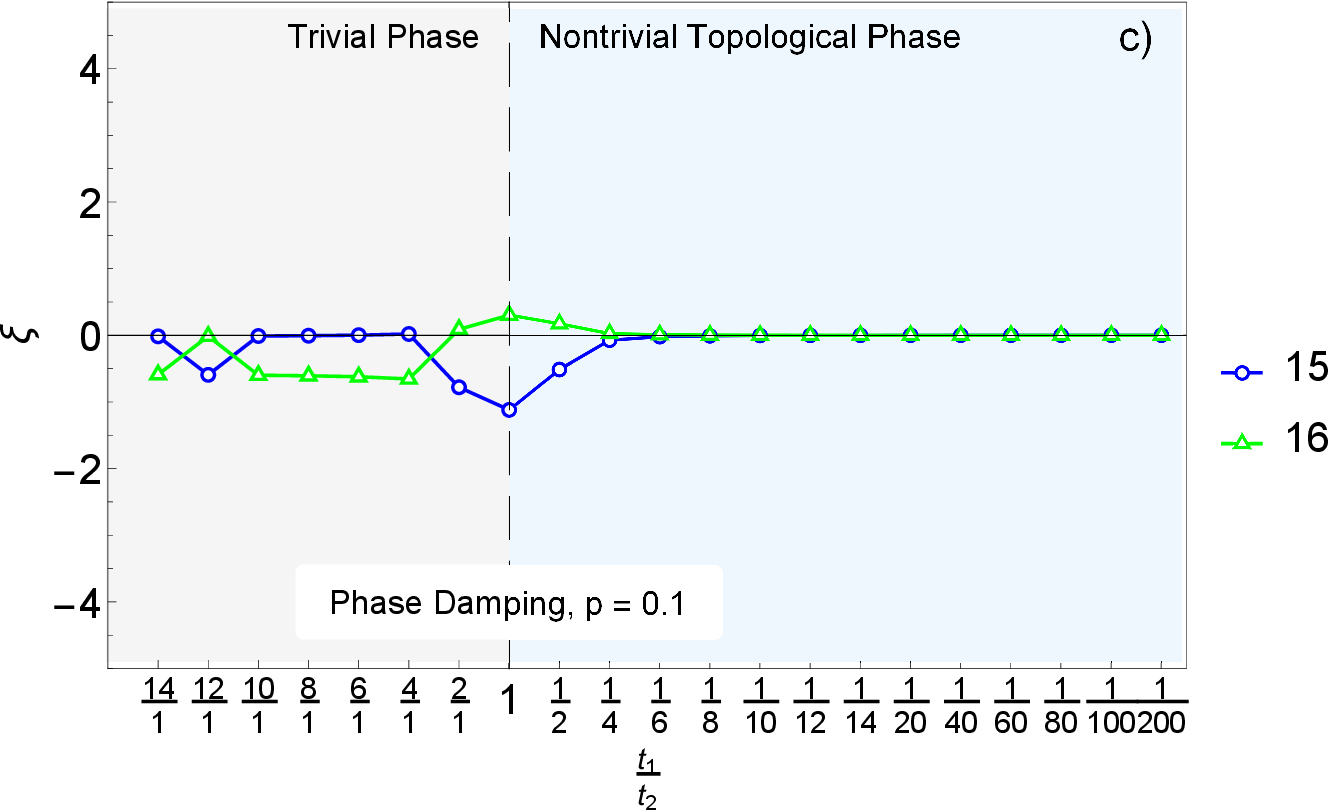}\quad
	\includegraphics[width=0.45\linewidth]{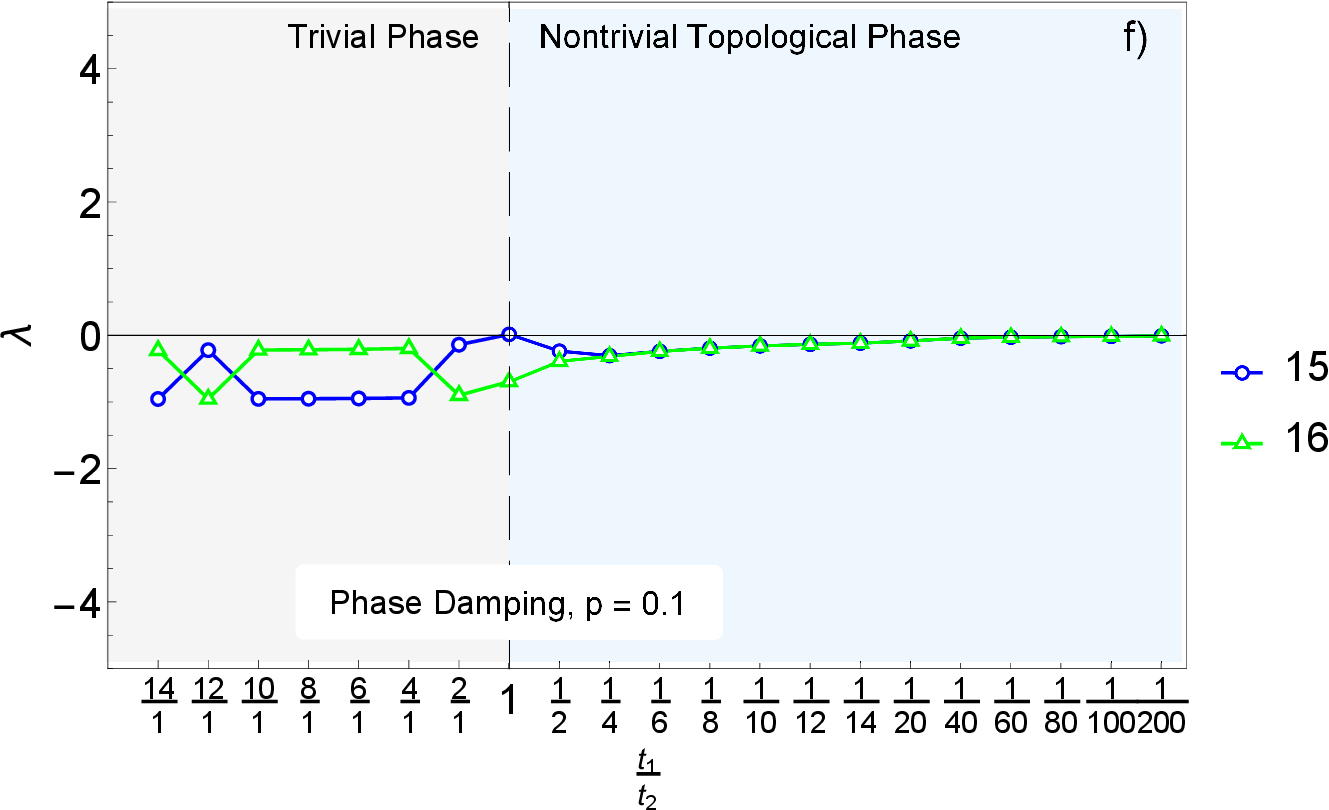} \\
	\caption{Effect of phase damping on the sixteen SSH fuels. The squeezing indicator \(\xi\) (left column) and displacement indicator \(\lambda\) (right column) are shown as functions of the dimerization ratio \(t_1/t_2\) for phase damping with \(p=0.1\), applied independently to each qubit of the four-qubit fuel. The fourteen bulk fuels retain finite coherent response, whereas the two edge fuels (states 15 and 16) remain at \(\lambda \approx \xi \approx 0\) throughout the topological regime \((t_1<t_2)\), preserving the pure-thermalization signature.\label{fig:PhaseDamping}}
\end{figure}

To test robustness against pure dephasing during flight, we apply an identical single-qubit phase-damping channel independently to each qubit of the four-qubit fuel. Its Kraus operators are
\begin{equation}\label{eq:DephasingKraus}
	K^{\mathrm{Deph}}_1=\sqrt{1-p}\,I, \qquad K^{\mathrm{Deph}}_2=\sqrt{p}\,Z,
\end{equation}
where \(p\) is the dephasing strength. Figure~\ref{fig:PhaseDamping} shows the resulting squeezing and displacement indicators for \(p=0.1\). The bulk fuels continue to exhibit finite coherent response, whereas the two edge fuels remain at \(\lambda \approx \xi \approx 0\) throughout the topological regime \((t_1<t_2)\), indicating that the edge-fuel pure-thermalization response is robust to phase damping at this level.

\newpage

%\clearpage \newpage
\subsection*{S3.b. Depolarizing channel}

\begin{figure}[b!]
	\includegraphics[width=0.47\linewidth]{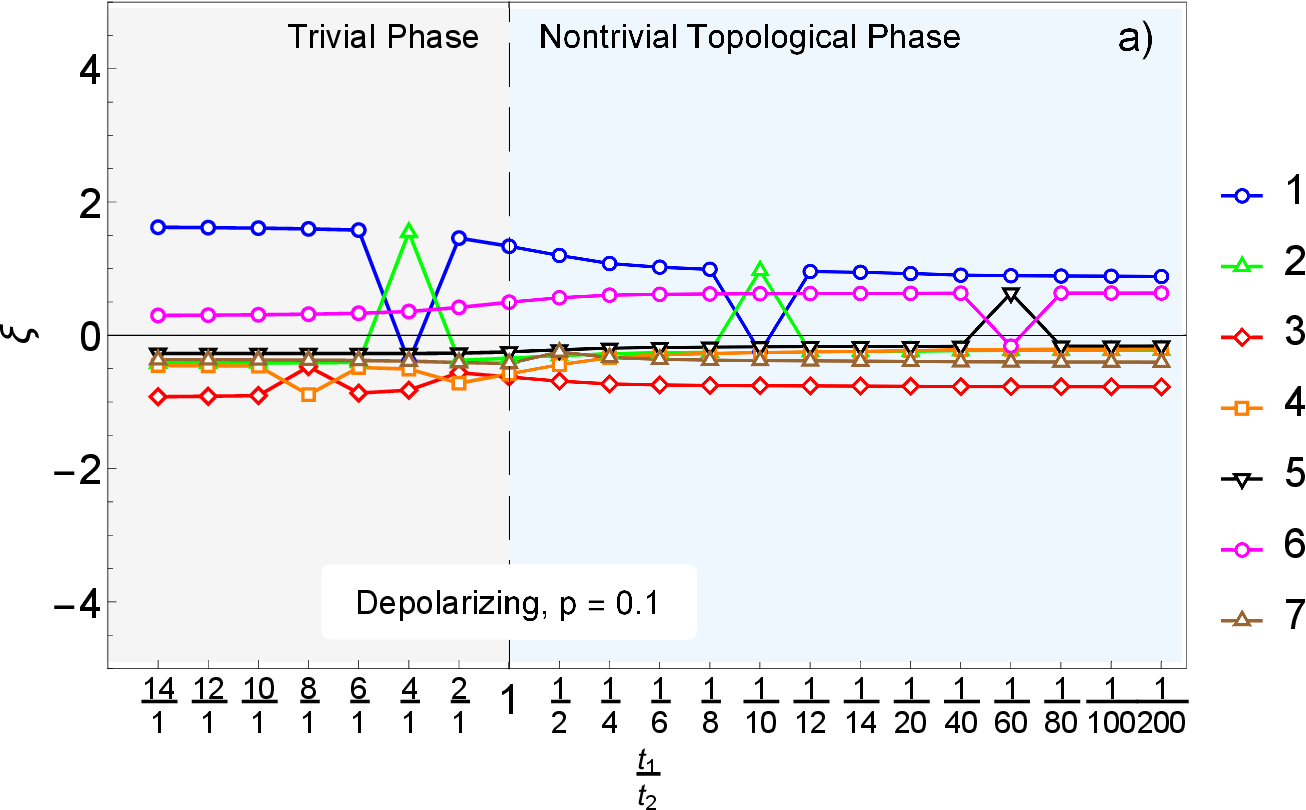}\quad
	\includegraphics[width=0.47\linewidth]{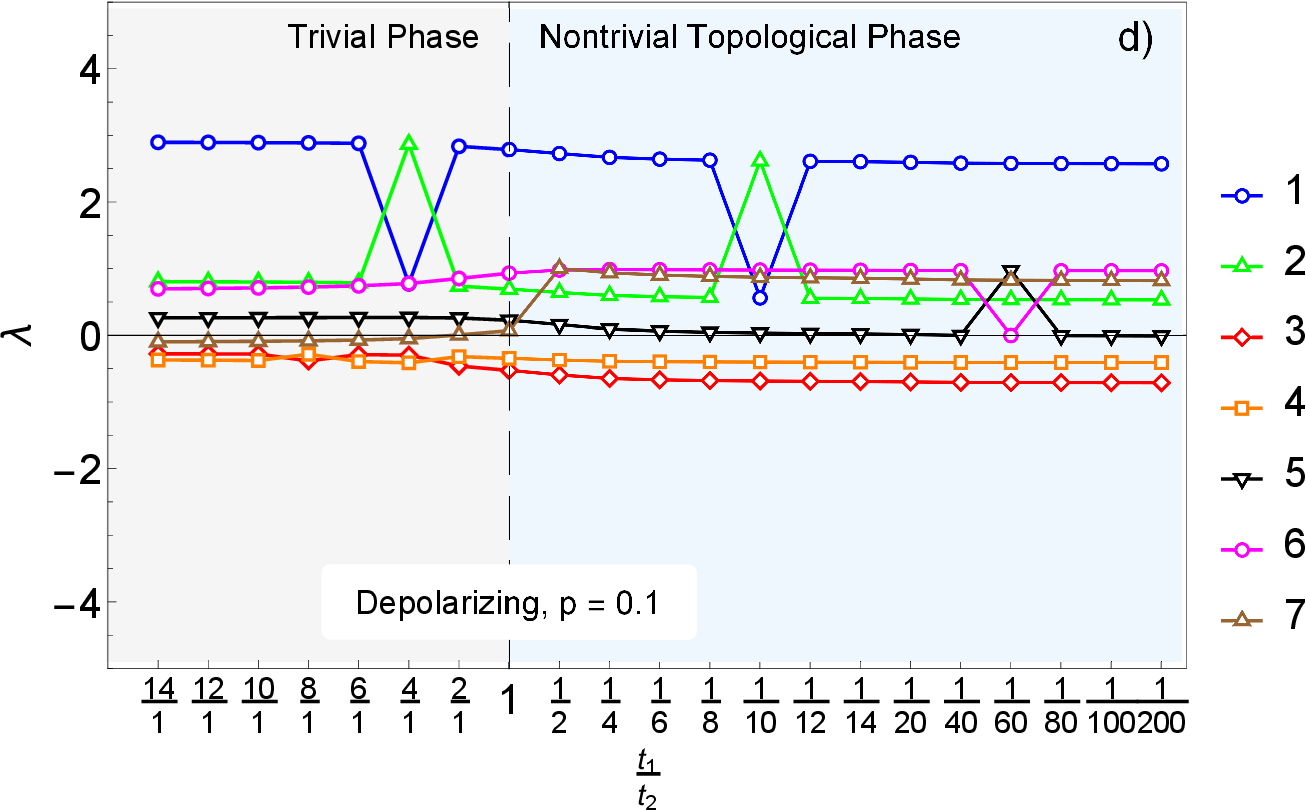} \\ \ \\
	\includegraphics[width=0.47\linewidth]{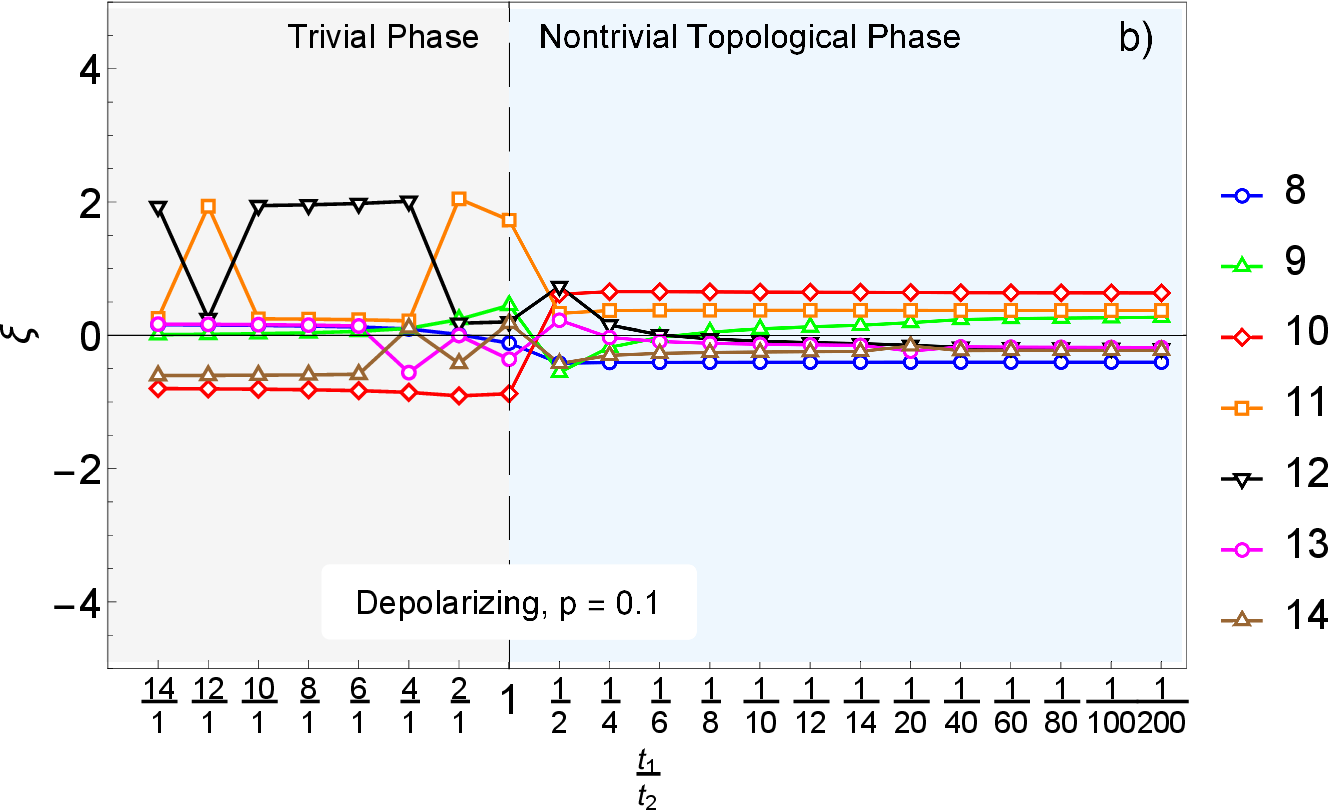}\quad
	\includegraphics[width=0.47\linewidth]{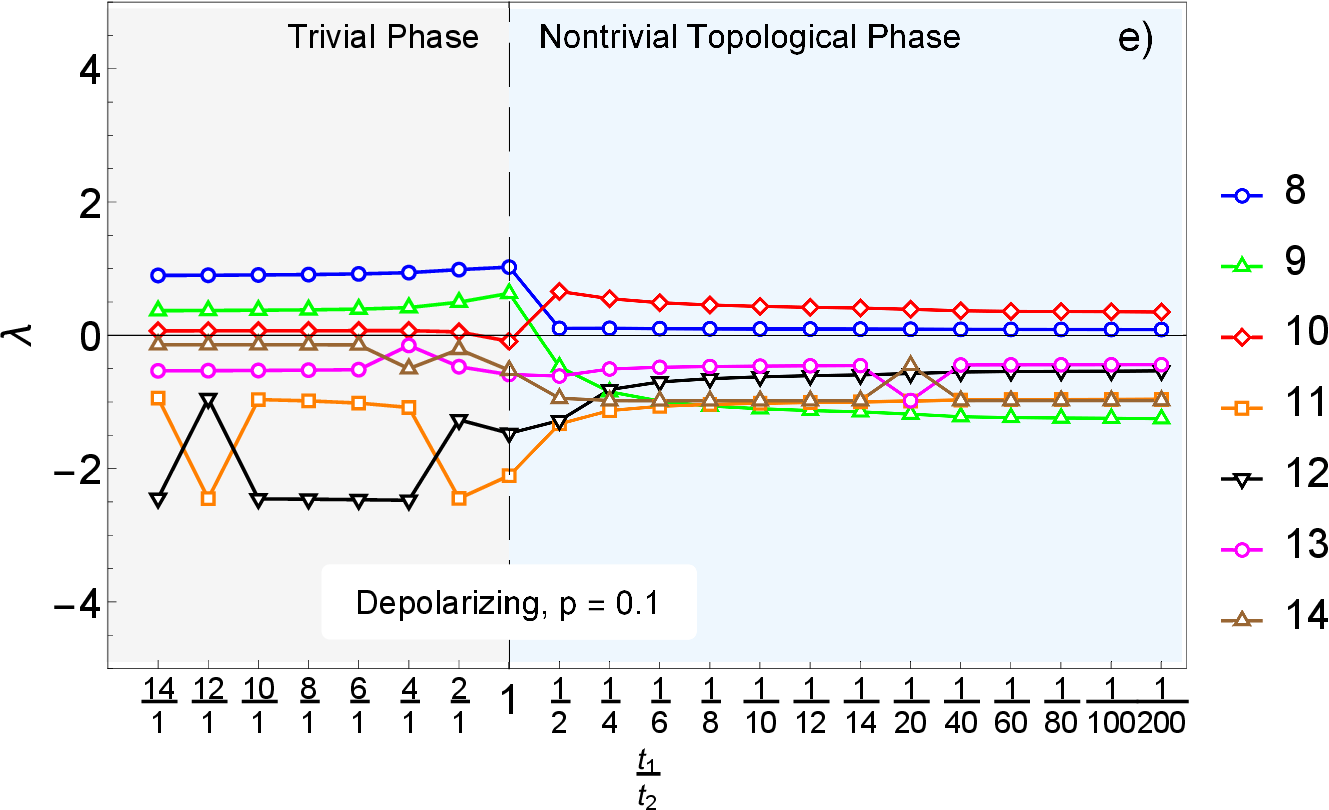} \\ \ \\
	\includegraphics[width=0.47\linewidth]{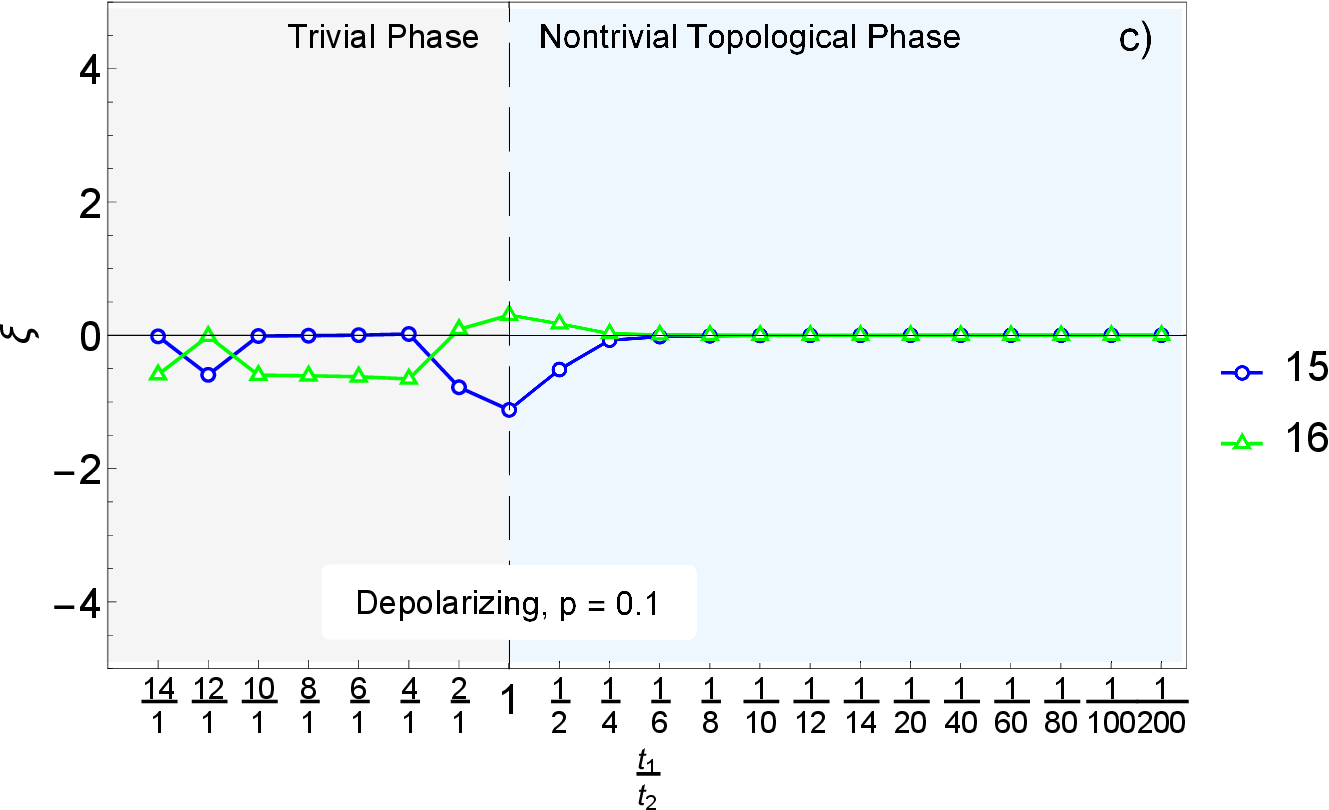}\quad
	\includegraphics[width=0.47\linewidth]{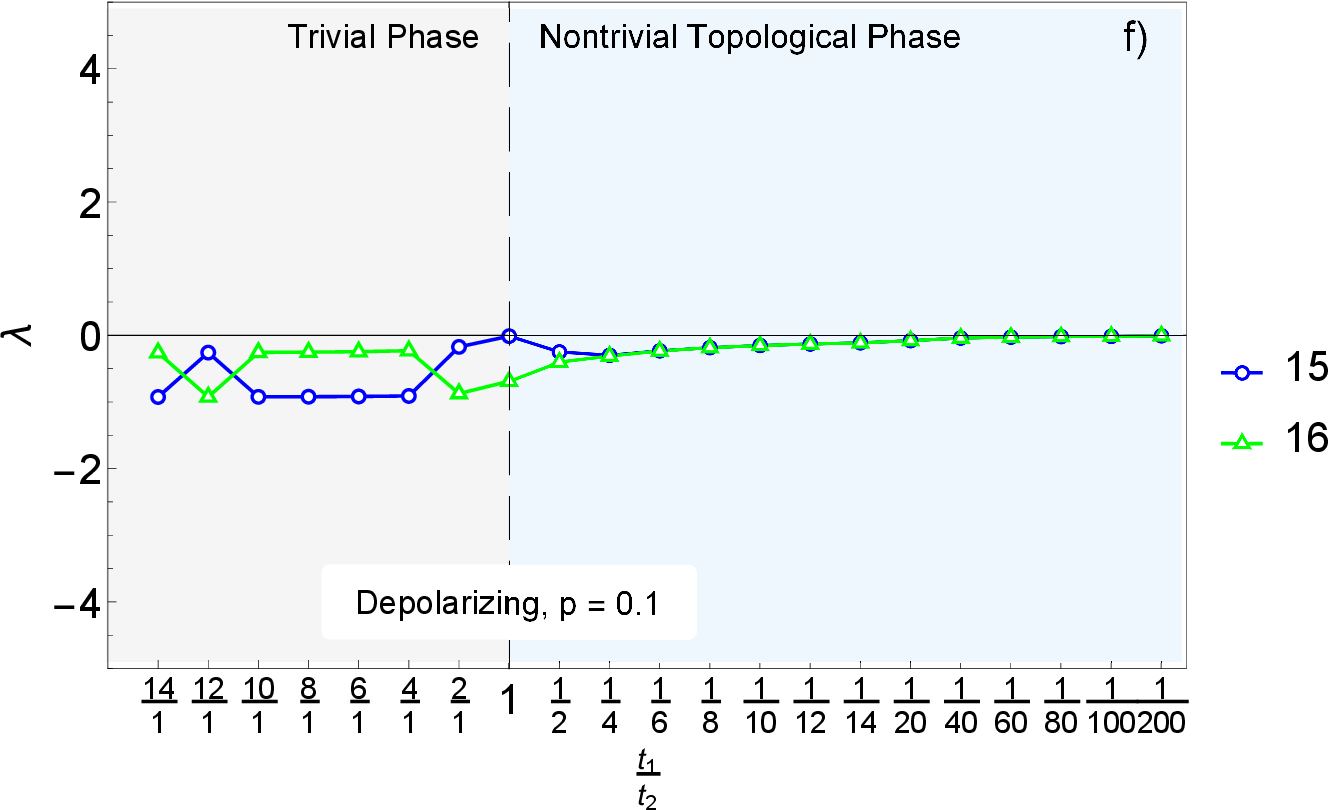} \\
	\caption{Effect of depolarizing noise on the sixteen SSH fuels. The squeezing indicator \(\xi\) (left column) and displacement indicator \(\lambda\) (right column) are shown as functions of the dimerization ratio \(t_1/t_2\) for depolarizing strength \(q=0.1\), applied independently to each qubit of the four-qubit fuel. The bulk fuels continue to exhibit finite coherent channels, whereas the two edge fuels (states 15 and 16) remain at \(\lambda \approx \xi \approx 0\) in the topological regime \((t_1<t_2)\), consistent with robust pure thermalization.\label{fig:Depolarizing}}
\end{figure}

We next consider single-qubit depolarization applied independently to each qubit of the four-qubit fuel. The Kraus operators are
\begin{equation}\label{eq:DepolarizationKraus} 
	K^{\mathrm{Depol}}_1=\sqrt{1-\frac{3q}{4}}\,I,\qquad
	K^{\mathrm{Depol}}_2=\frac{\sqrt{q}}{2}X,\qquad
	K^{\mathrm{Depol}}_3=\frac{\sqrt{q}}{2}Y,\qquad
	K^{\mathrm{Depol}}_4=\frac{\sqrt{q}}{2}Z,
\end{equation}
where \(q\) denotes the depolarizing strength. Fig.~\ref{fig:Depolarizing} presents the corresponding results for \(q=0.1\). As in the phase-damping case, the two edge fuels remain in the pure-thermalization sector with \(\lambda \approx \xi \approx 0\) across the topological regime, whereas the bulk fuels retain finite coherent channels.

\clearpage

\newpage
\section{S4. ROBUSTNESS OF THE EDGE-FUEL RESPONSE UNDER BOND DISORDER AND MODERATE ONSITE DISORDER}\label{sec:S4}
\subsection{S4.a. Bond disorder}
To test robustness against static bond disorder, we perturb each hopping amplitude of the SSH Hamiltonian independently by a random amount drawn uniformly up to \(5\%\) of its nominal value. Figure~S6 shows the resulting squeezing and displacement indicators. The bulk fuels continue to exhibit finite coherent response, whereas the two edge fuels remain at \(\lambda \approx \xi \approx 0\) throughout the topological regime \((t_1<t_2)\), indicating that the edge-fuel pure-thermalization signature is robust to bond disorder at this level.

\begin{figure}[h!]
	\includegraphics[width=0.47\linewidth]{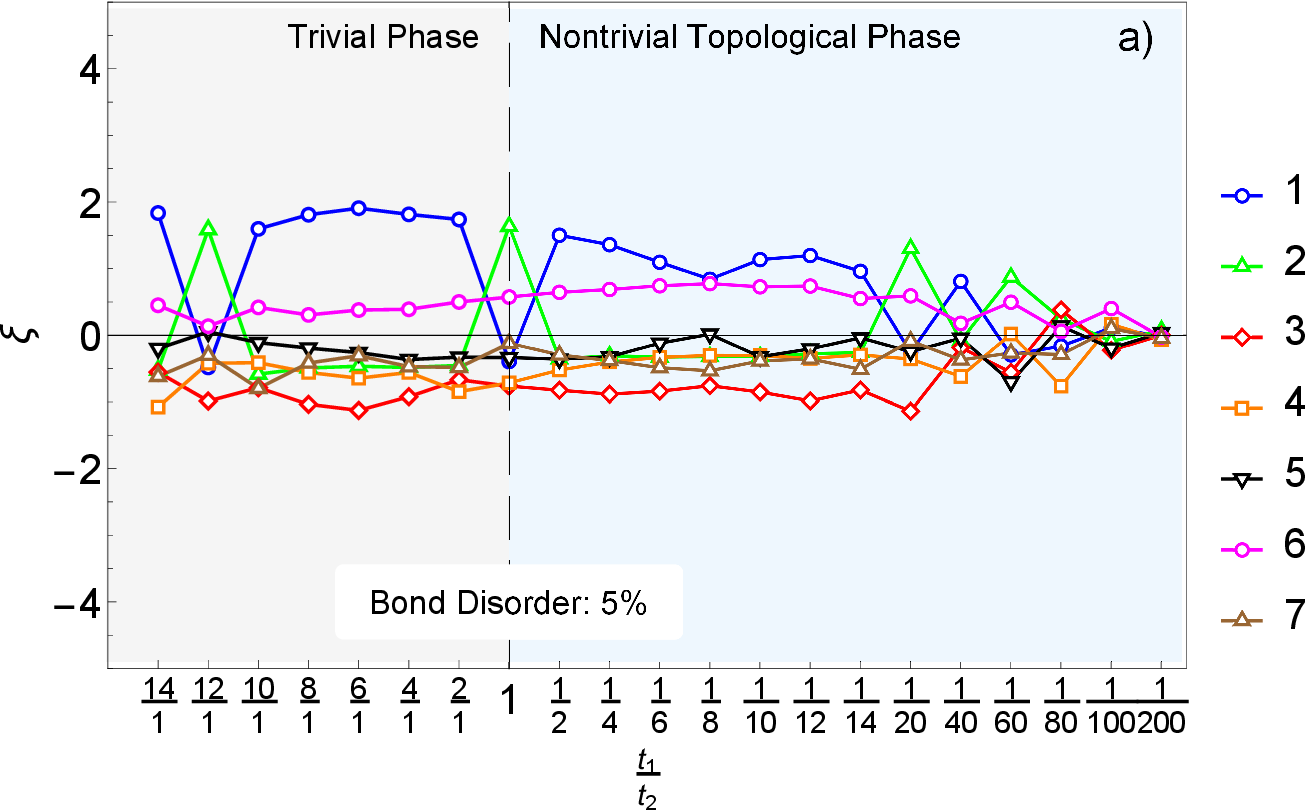}\quad
	\includegraphics[width=0.47\linewidth]{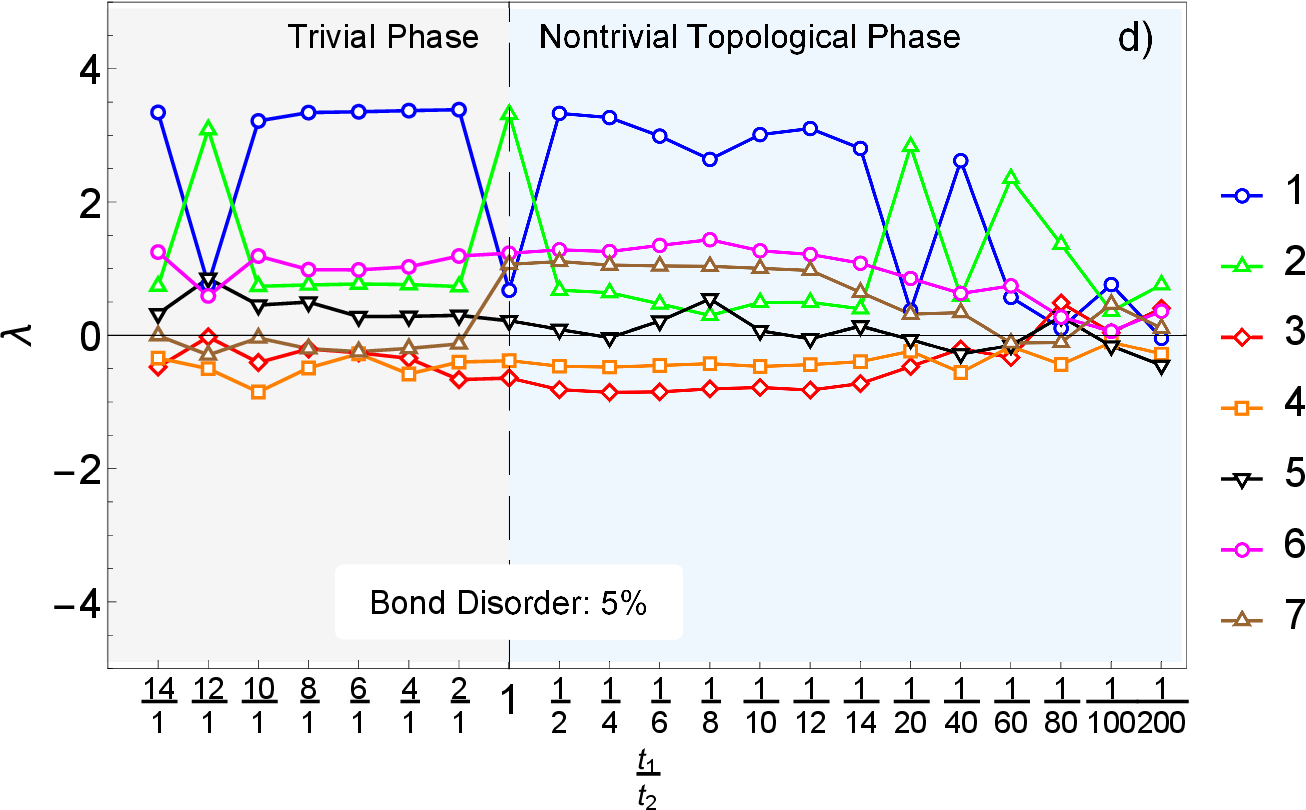} \\ \ \\
	\includegraphics[width=0.47\linewidth]{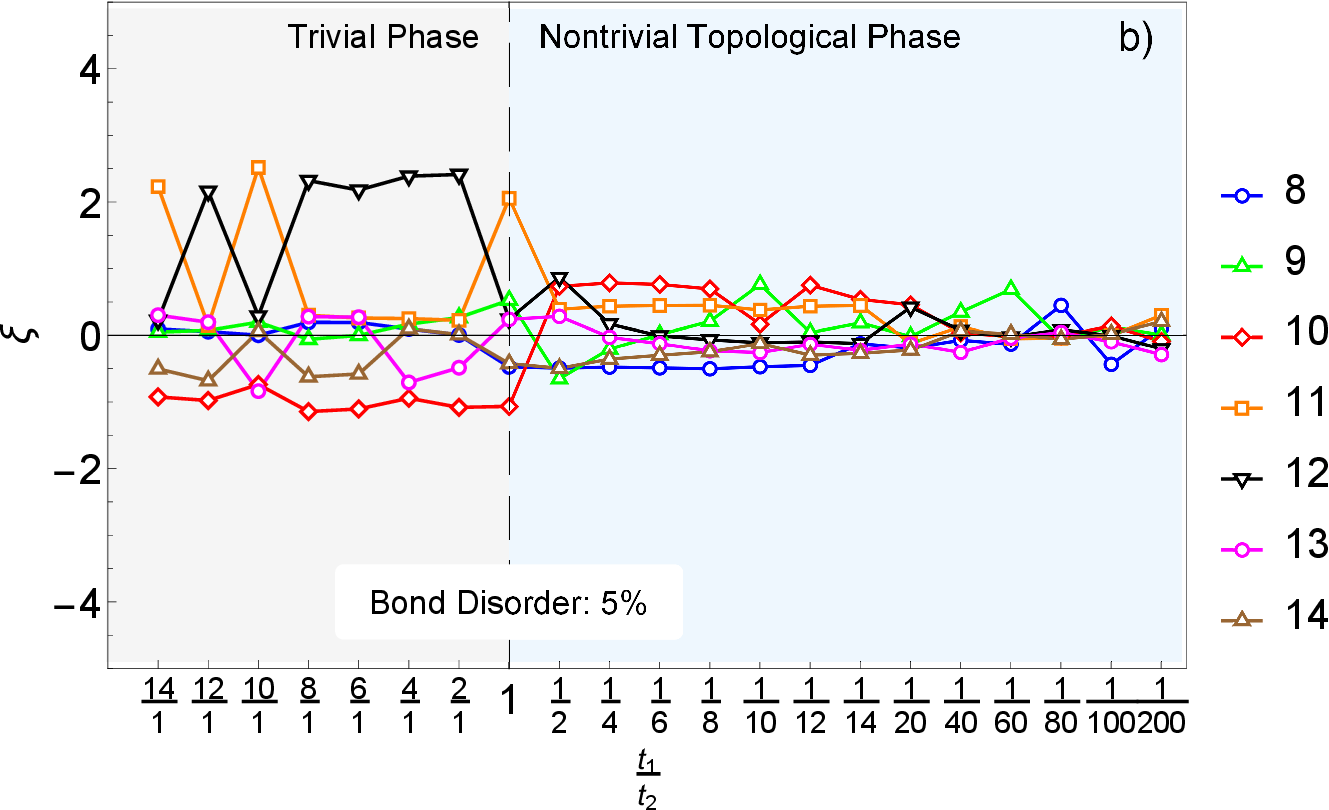}\quad
	\includegraphics[width=0.47\linewidth]{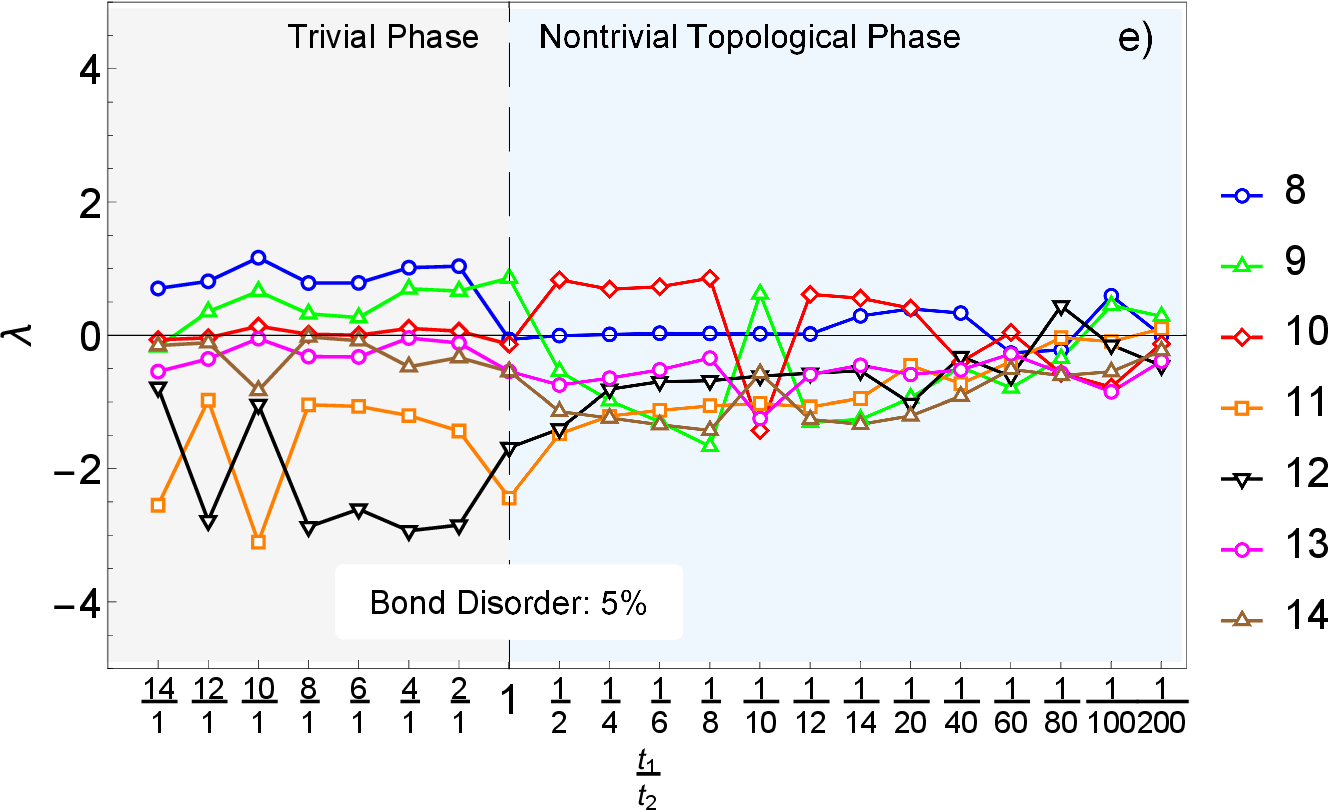} \\ \ \\
	\includegraphics[width=0.47\linewidth]{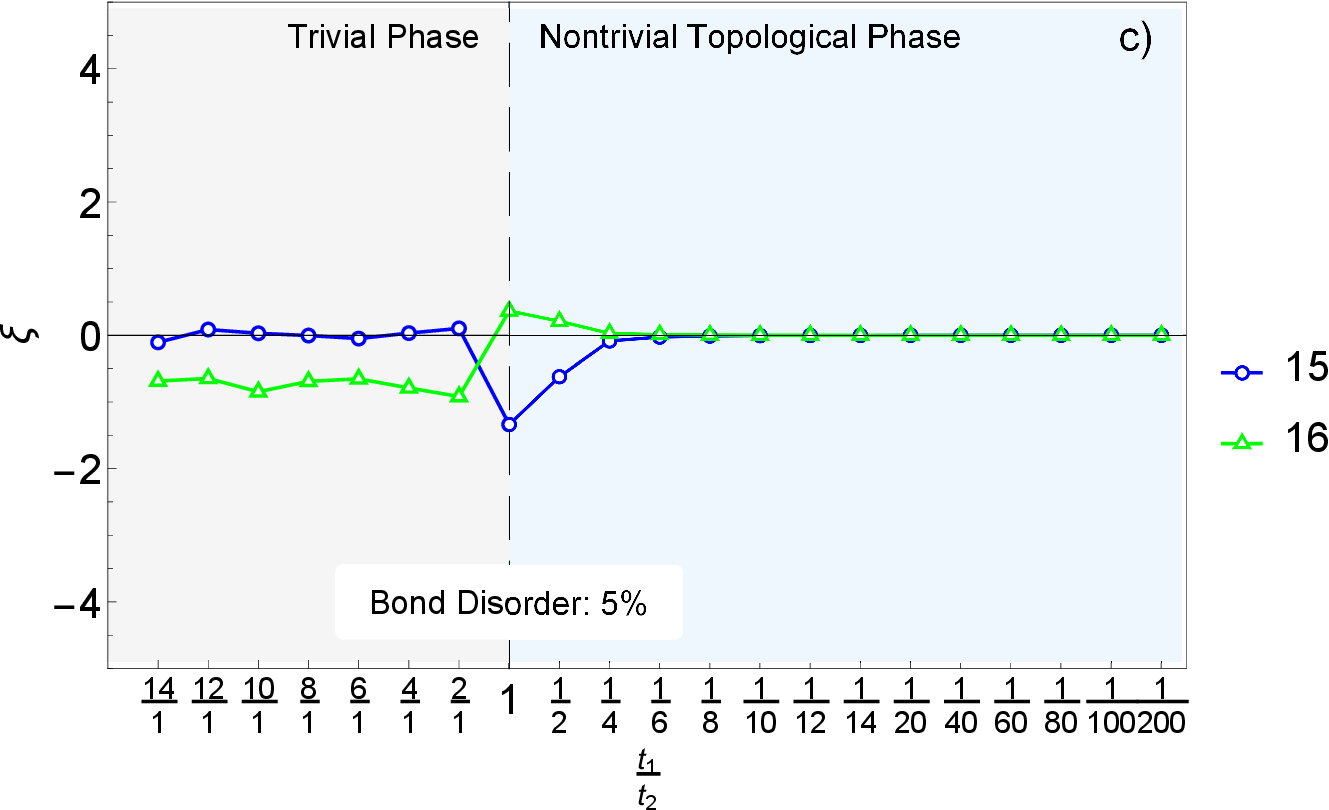}\quad
	\includegraphics[width=0.47\linewidth]{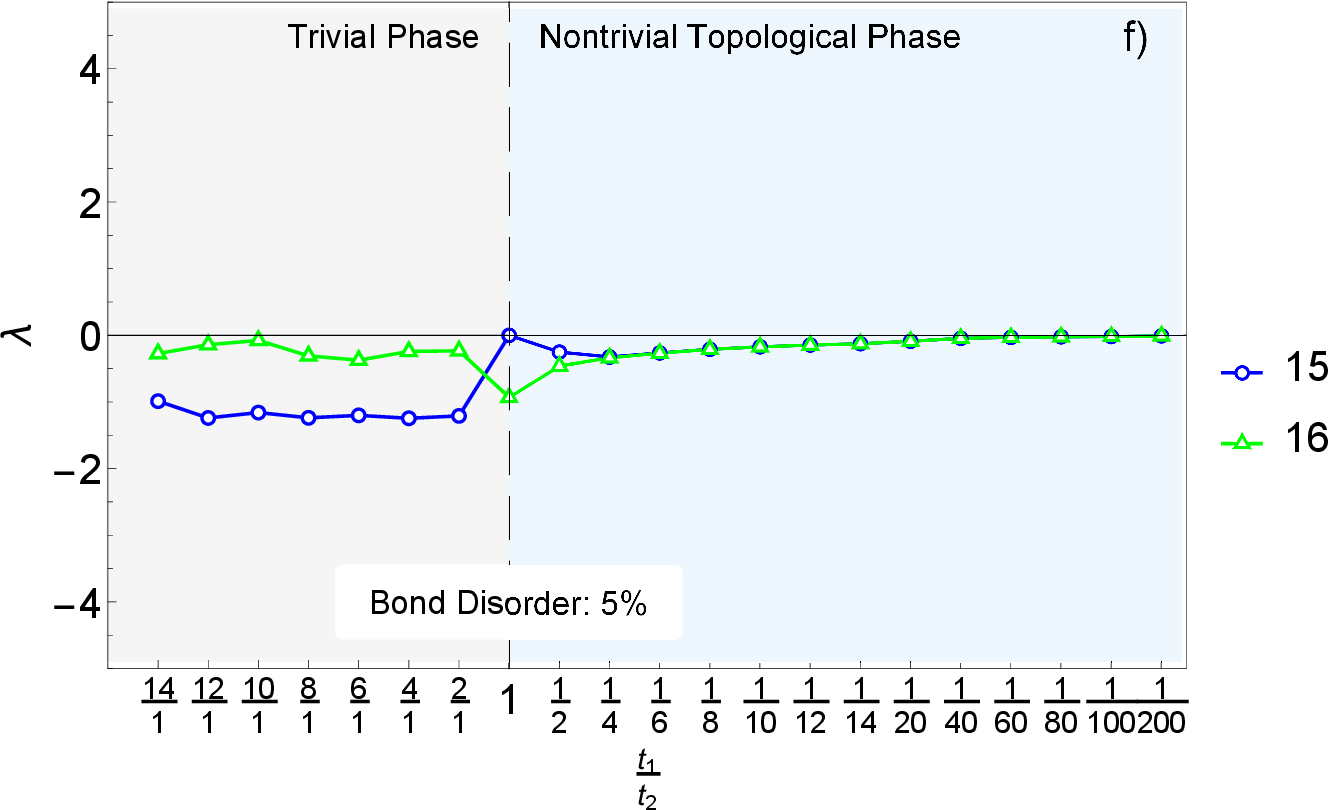} \\
	\caption{Effect of bond disorder on the sixteen SSH fuels. The squeezing indicator \(\xi\) (left column) and displacement indicator \(\lambda\) (right column) are shown as functions of the dimerization ratio \(t_1/t_2\) after adding independent random perturbations of up to \(5\%\) to each hopping amplitude of the SSH Hamiltonian. The fourteen bulk fuels retain finite coherent response, whereas the two edge fuels (states 15 and 16) remain at \(\lambda \approx \xi \approx 0\) throughout the topological regime \((t_1<t_2)\), preserving the pure-thermalization signature.
		\label{fig:bonddisorder}}
\end{figure} 
%\end{widetext}
%\twocolumngrid

\newpage
\subsection{S4.b. Onsite disorder}
\begin{figure}[h!]
	\includegraphics[width=0.47\linewidth]{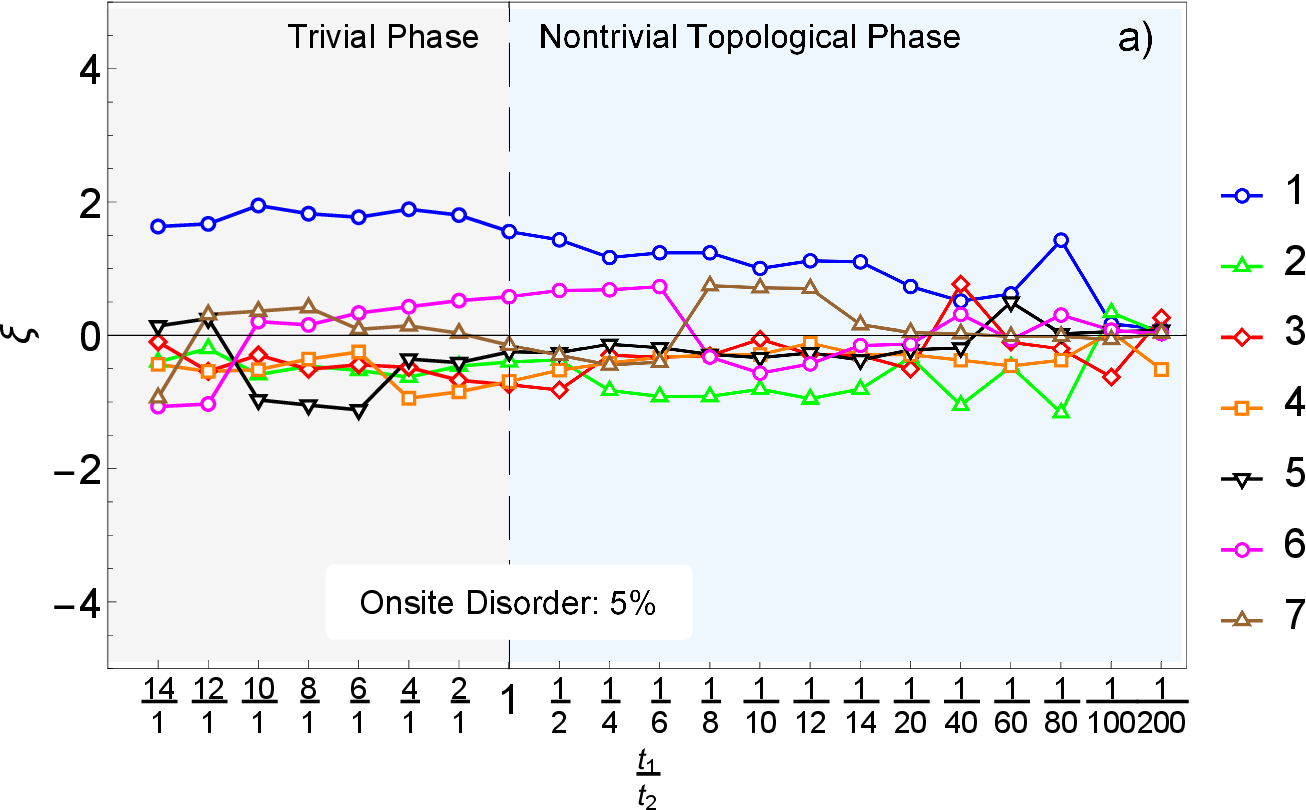}\quad
	\includegraphics[width=0.47\linewidth]{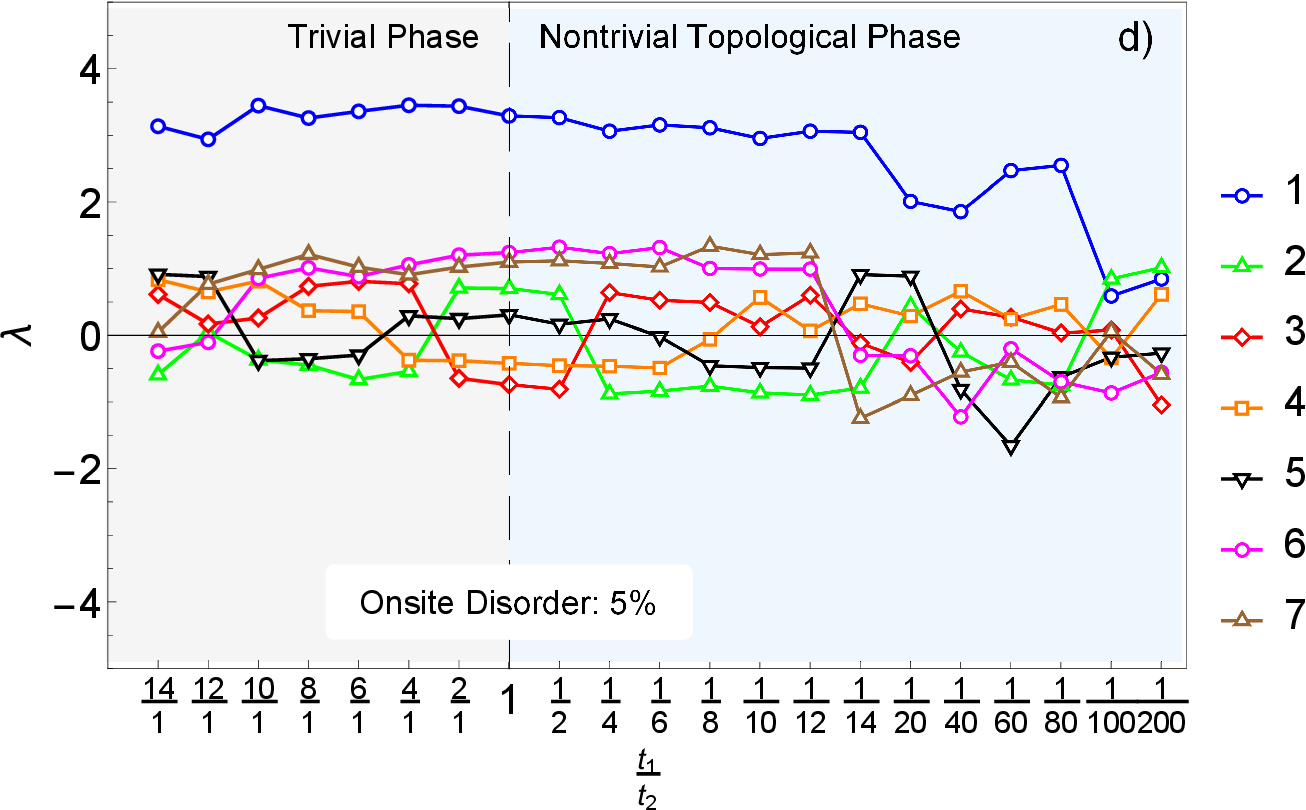} \\ \ \\
	\includegraphics[width=0.47\linewidth]{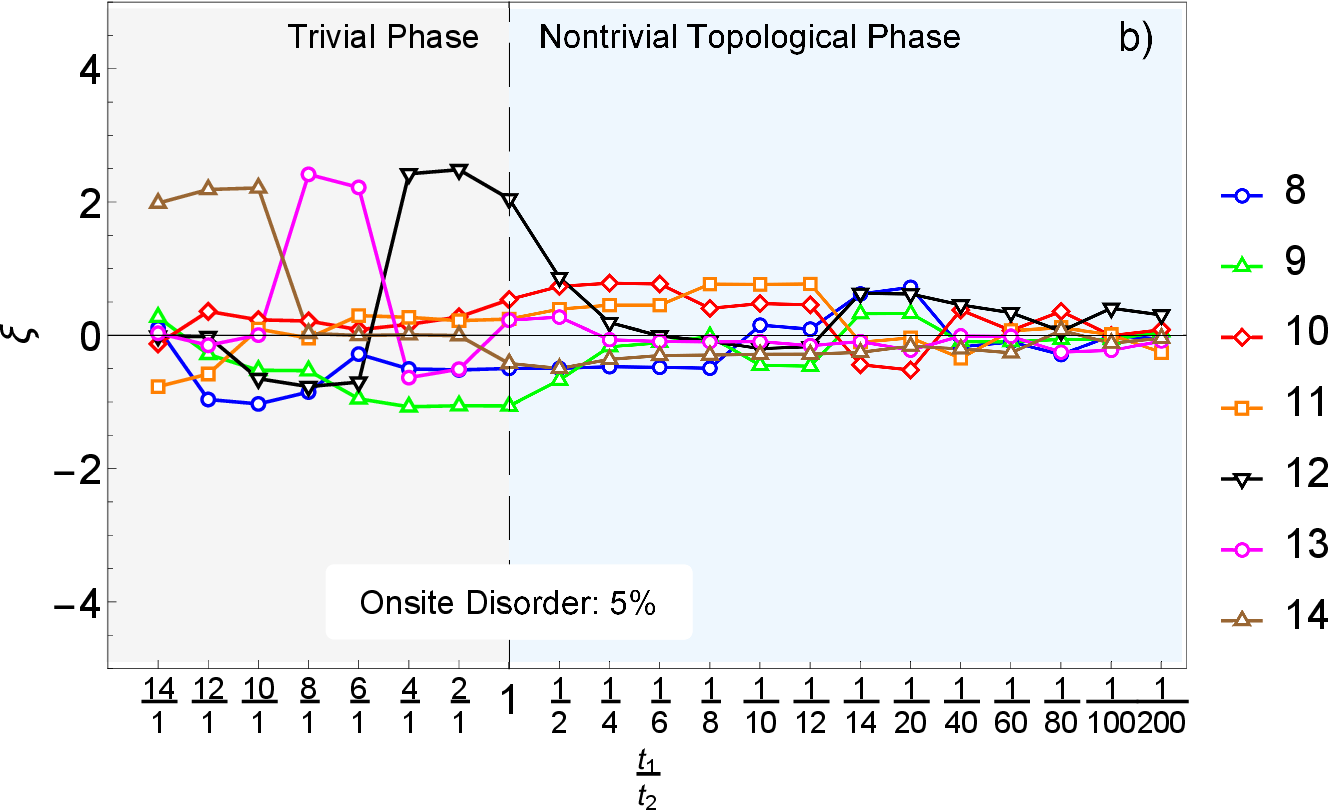}\quad
	\includegraphics[width=0.47\linewidth]{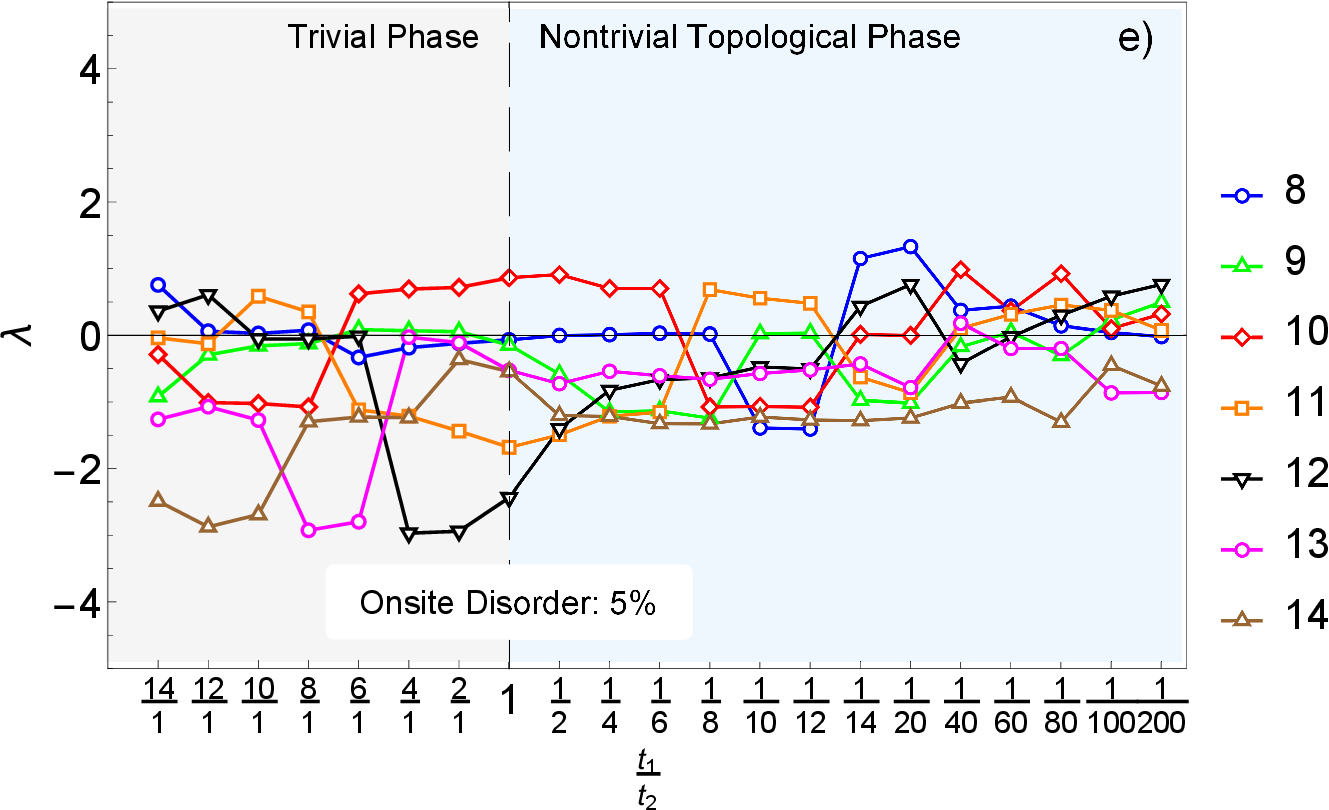} \\ \ \\
	\includegraphics[width=0.47\linewidth]{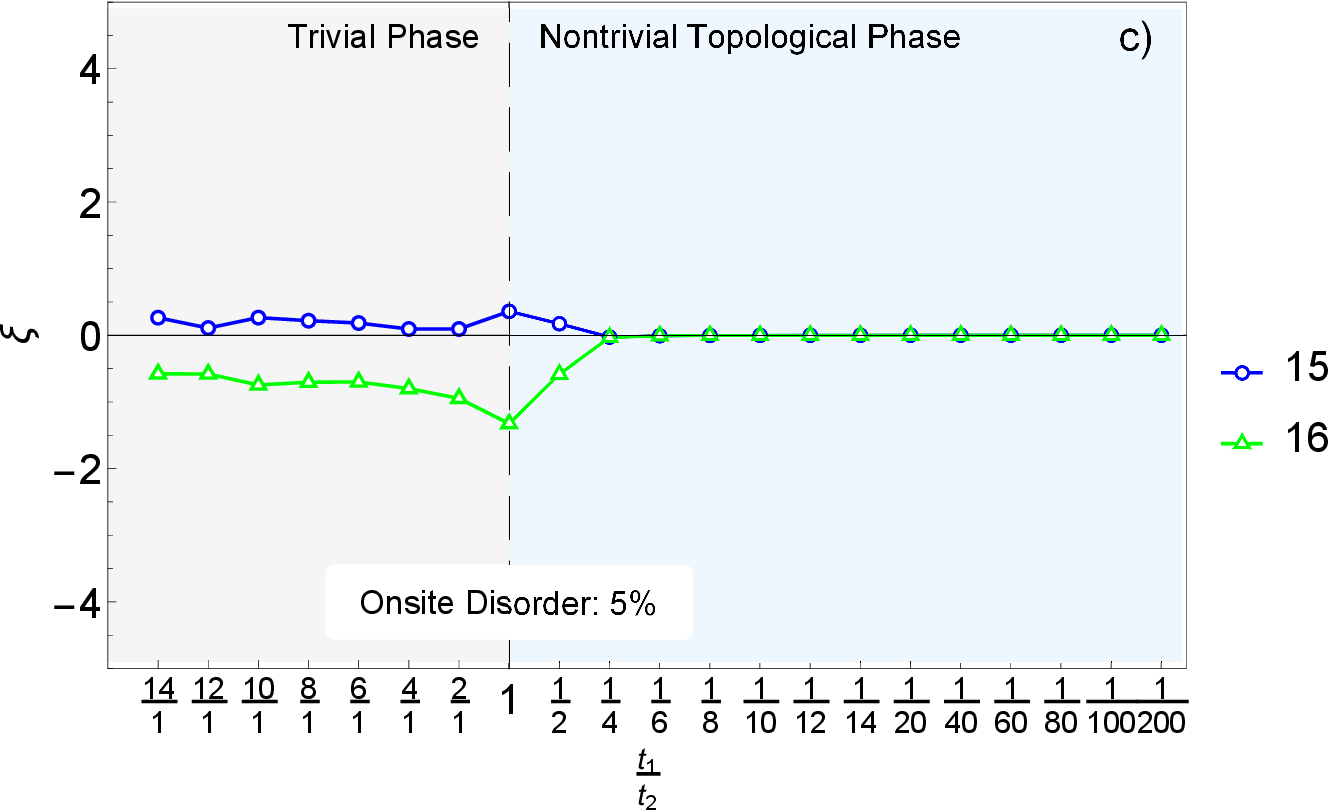}\quad
	\includegraphics[width=0.47\linewidth]{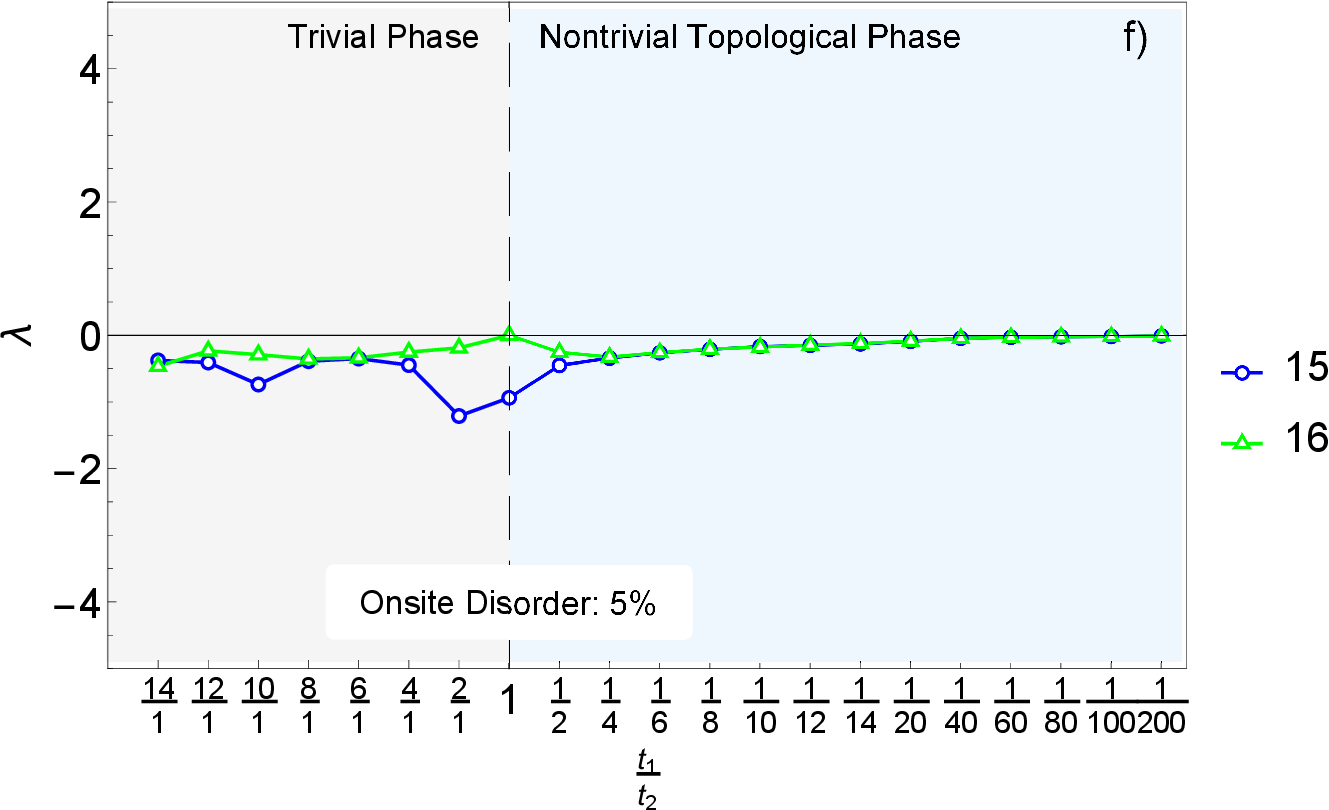} \\
	\caption{Effect of onsite disorder on the sixteen SSH fuels. The squeezing indicator \(\xi\) (left column) and displacement indicator \(\lambda\) (right column) are shown as functions of the dimerization ratio \(t_1/t_2\) after adding independent onsite perturbations to the diagonal entries of the SSH Hamiltonian, each drawn up to \(5\%\) of the largest matrix-element magnitude. The bulk fuels continue to exhibit finite coherent channels, whereas the two edge fuels (states 15 and 16) remain at \(\lambda \approx \xi \approx 0\) in the topological regime \((t_1<t_2)\), consistent with robust pure thermalization.\label{fig:onsitedisorder}}
\end{figure}

	We next consider \emph{moderate} onsite disorder by adding independent random perturbations to the diagonal entries of the SSH Hamiltonian, each drawn uniformly with magnitude up to $5\%$ of the largest matrix-element magnitude. Because generic onsite disorder breaks the chiral sublattice symmetry of the SSH model, we do not claim exact symmetry-protected robustness under arbitrary onsite perturbations. Rather, we test a moderate-disorder regime in which the finite-chain edge-localized modes and their operational thermodynamic signature remain clearly identifiable. Figure~\ref{fig:onsitedisorder} presents the corresponding results. As in the bond-disorder case, the two edge fuels remain in the pure-thermalization sector with $\lambda \approx \xi \approx 0$ across the topological regime, whereas the bulk fuels continue to show finite coherent channels.

\newpage
\section{S5. Absence of an analogous pure-thermalization regime in representative trivial lattice models}\label{sec:S5}
In this section we test whether the pure-thermalization behaviour found for the SSH edge fuels also arises in representative lattice models without nontrivial topology. Our aim is not to establish a no-go theorem, but to assess, through explicit examples and broad numerical sweeps, whether an analogous regime with simultaneously vanishing displacement and squeezing indicators can be found in simple trivial settings. For each model, we evaluated the same micromaser indicators \(\lambda\) and \(\xi\) used in the main text over a substantially wider set of parameters that can be displayed compactly here. The figures below therefore show representative parameter sweeps from this broader search. Across the cases explored, we did not find an analogous pure-thermalization regime.

\begin{figure}[h!]
	\includegraphics[width=0.4\linewidth]{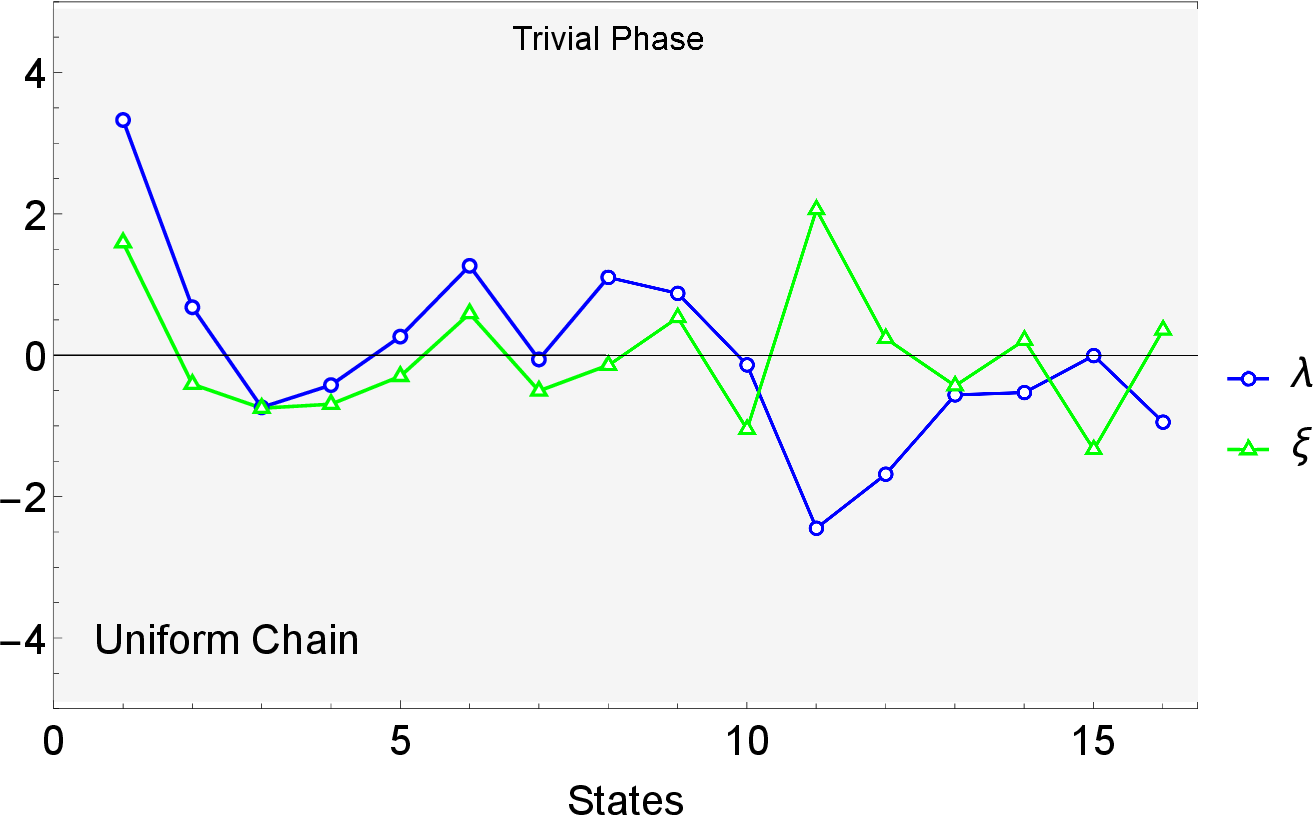}\\ \ \\ \ \\
	\includegraphics[width=0.4\linewidth]{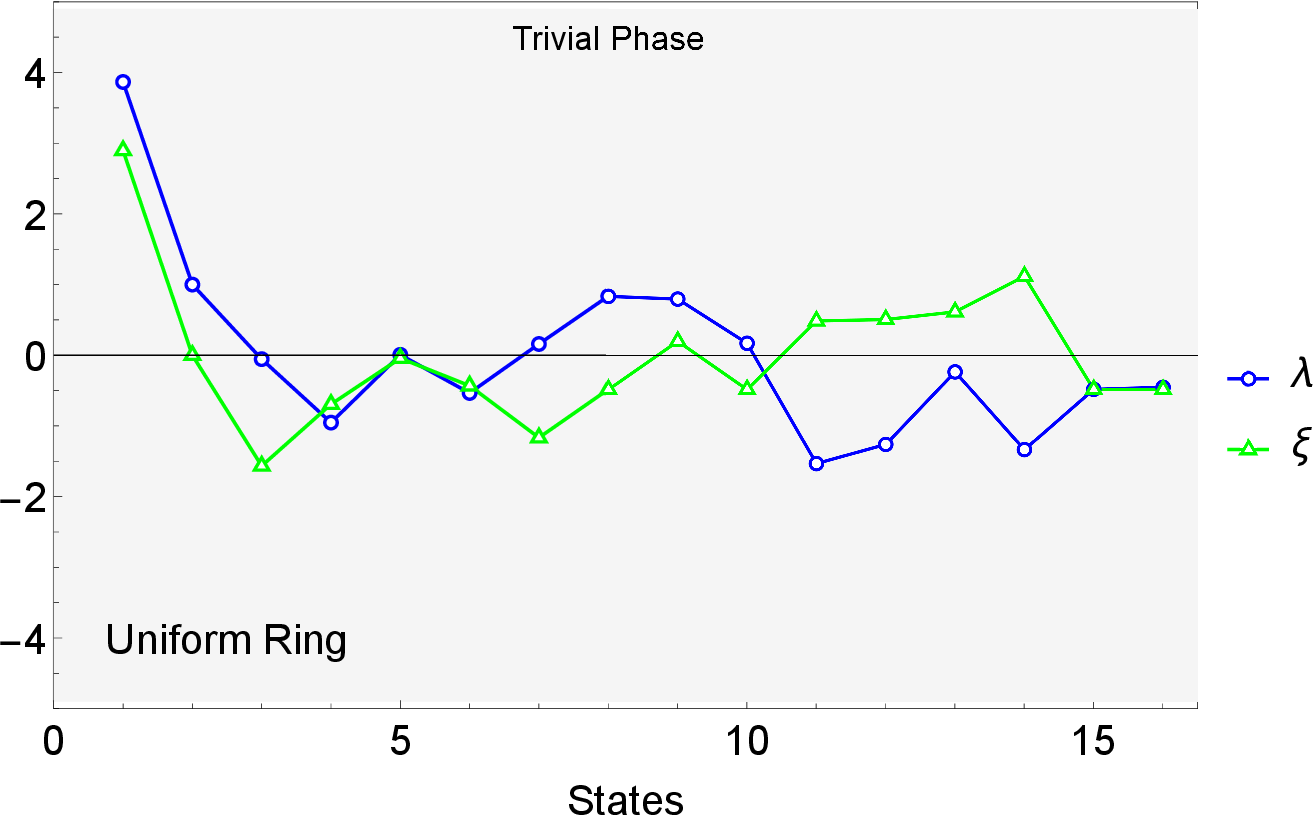}\quad
	\includegraphics[width=0.4\linewidth]{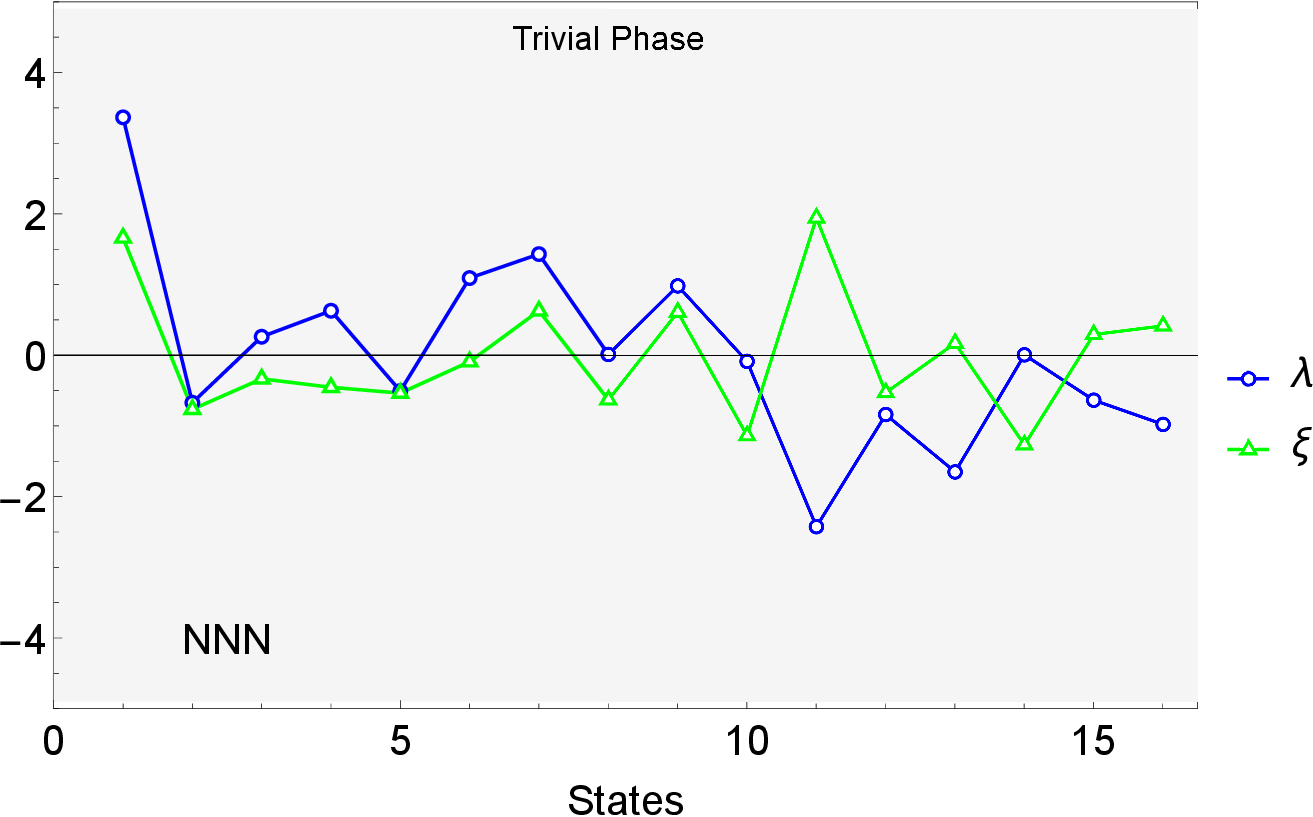}\\ \ \\ \ \\
	\includegraphics[width=0.4\linewidth]{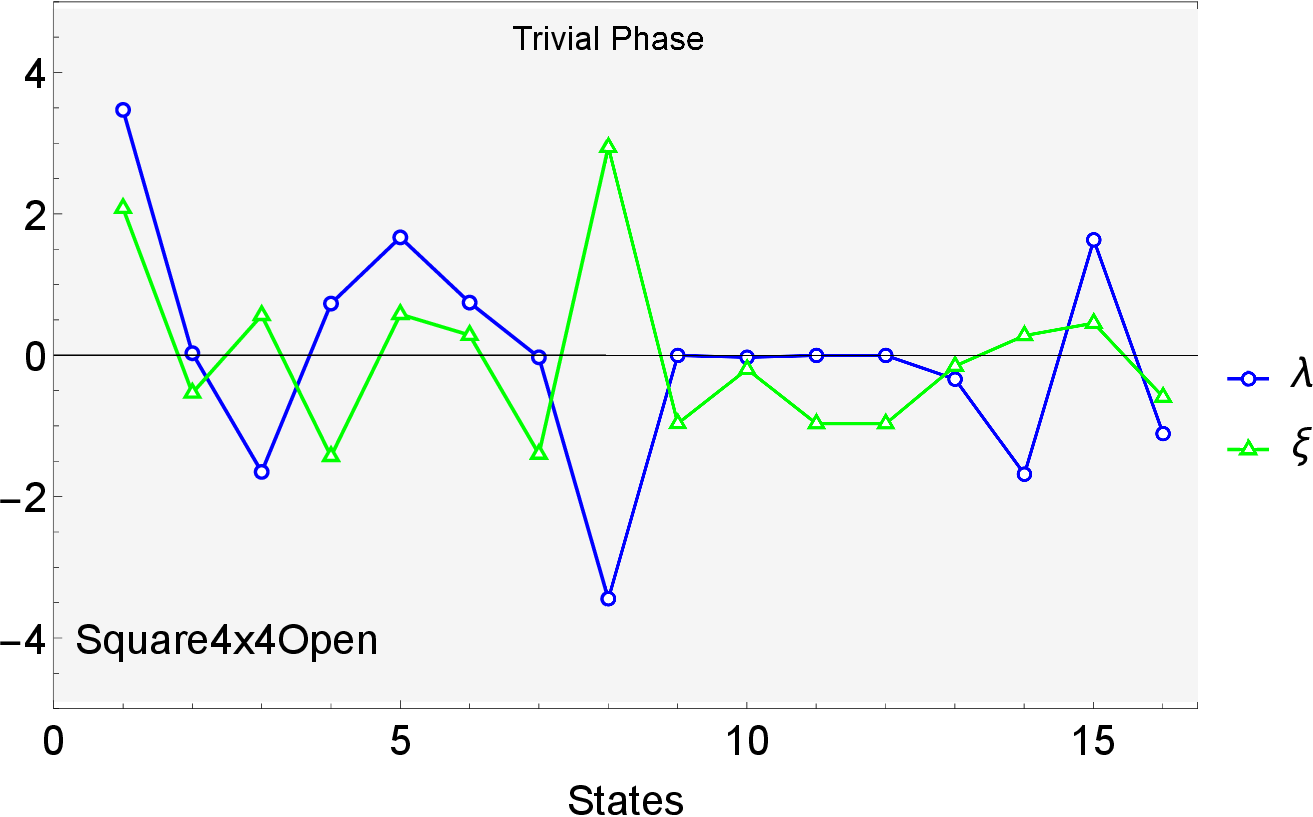}\quad
	\includegraphics[width=0.4\linewidth]{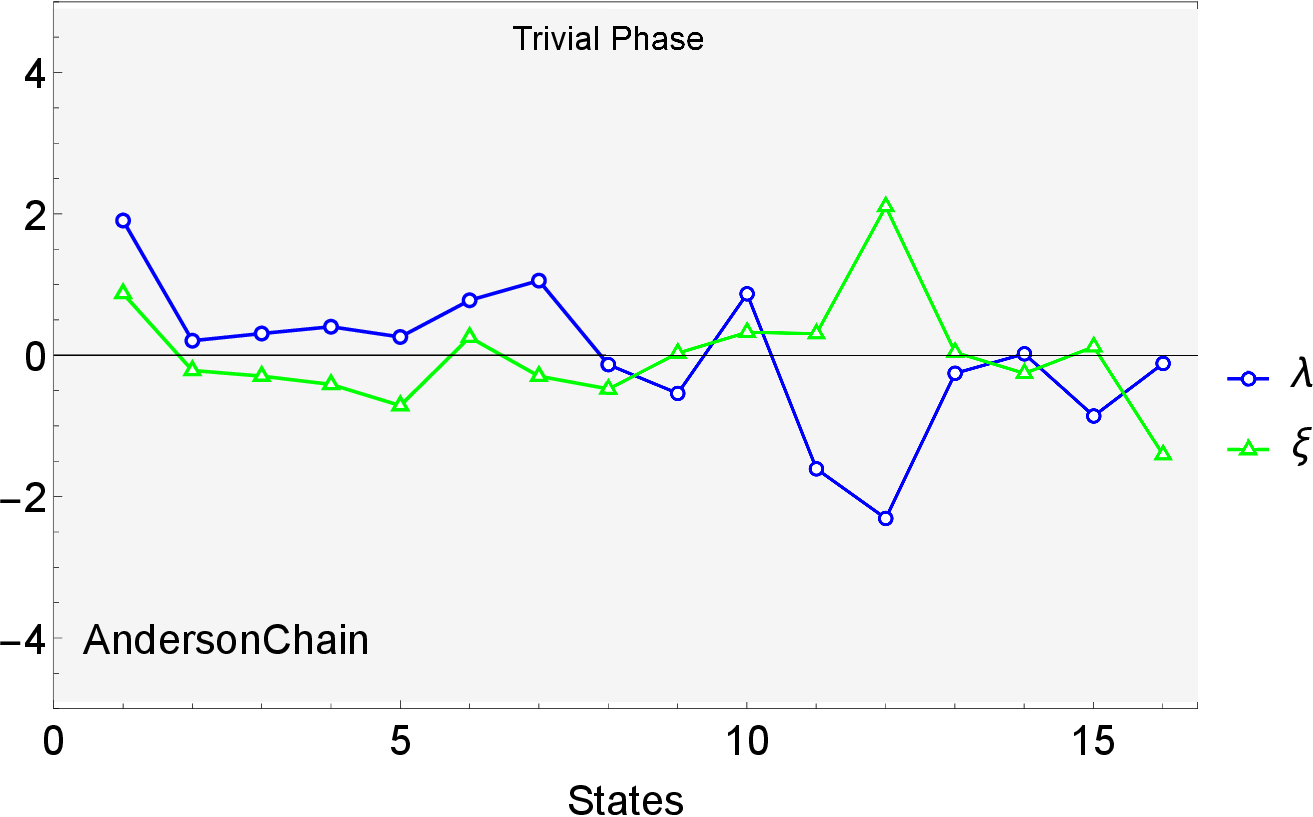}	
	\caption{Representative parameter sweeps for the trivial models. The squeezing indicator \(\xi\) and displacement indicator \(\lambda\) are evaluated for fuels constructed from eigenstates of the trivial models. The panels shown are representative of a broader numerical search and do not reveal an extended regime with simultaneous suppression of both coherent indicators analogous to the SSH edge-fuel case.
	\label{fig:figLattice1}}
\end{figure}

As representative trivial model controls, we consider a uniform 1D tight-binding chain with open boundaries,
\begin{equation}
	H_{\mathrm{uni}} \;=\; t \sum_{j=1}^{N-1} \bigl( \,|j\rangle\langle j{+}1| \;+\; |j{+}1\rangle\langle j| \,\bigr),
\end{equation}

\noindent
a uniform 1D ring,
\begin{equation}
	H_{\mathrm{ring}} \;=\; t \sum_{j=1}^{N} \Bigl(\, |j\rangle\langle j{+}1| \;+\; |j{+}1\rangle\langle j| \,\Bigr),
\end{equation}
a 1D chain with NNN (next nearest neighbor) interaction,
\begin{equation}
	H_{\mathrm{NNN}}
	= t_1 \sum_{j=1}^{N-1} \bigl( |j\rangle\langle j{+}1| + |j{+}1\rangle\langle j| \bigr)
	+ t_{nn} \sum_{j=1}^{N-2} \bigl( |j\rangle\langle j{+}2| + |j{+}2\rangle\langle j| \bigr),
\end{equation}
a 4×4 square lattice, open boundaries, NN hopping $t$
\begin{equation}
	H_{SL} \;=\; 
	t \sum_{x=1}^{3}\sum_{y=1}^{4} \bigl( |x,y\rangle\langle x{+}1,y| + |x{+}1,y\rangle\langle x,y| \bigr)
	\;+\;
	t \sum_{x=1}^{4}\sum_{y=1}^{3} \bigl( |x,y\rangle\langle x,y{+}1| + |x,y{+}1\rangle\langle x,y| \bigr),
\end{equation}
and a 1D Anderson chain (open boundaries): random onsite energies on the diagonal and uniform nearest-neighbor hopping
\begin{equation}
H_{\mathrm{And}} \;=\; \sum_{j=1}^{N} \varepsilon_j \, |j\rangle\langle j|
\;+\; t \sum_{j=1}^{N-1} \Bigl( |j\rangle\langle j{+}1| \;+\; |j{+}1\rangle\langle j| \Bigr),
\qquad (N=16,\ t=1),
\end{equation}
all of which are topologically trivial in the present context. Using their eigenstates as fuel states, we evaluated the same indicators \(\xi\) and \(\lambda\) over broad parameter sweeps. Figure~\ref{fig:figLattice1} shows representative subsets of these scans. Within the cases explored, we did not find an analogue of the SSH edge-fuel pure-thermalization regime.

%\newpage

\section*{S6. Density Matrix Elements Forming the Displacement and Squeezing Coherences}\label{sec:S6}

For four qubits the explicit  displacement and squeezing coherence terms summed by $\lambda$ and $\xi$, respectively, are given in Refs.~\cite{tuncer2019work,ozaydin2024engineering} as

\begin{eqnarray}
	\lambda &=& a_{1,2}+a_{1,3}+a_{1,4}+a_{1,5}+  a_{2,6}+a_{2,9}+a_{2,10}	\nonumber \\
	&+& a_{3,7} +a_{3,9}+a_{3,11}+a_{4,8}+a_{4,10}+a_{4,11}					\nonumber \\
	&+&a_{5,6}+a_{5,7}+a_{5,8}+ a_{6,13}+a_{6,14}+ a_{7,13}+a_{7,15}  		\nonumber \\
	&+&a_{8,14}+a_{8,15} + a_{9,12}+a_{9,13}+ a_{10,12}+a_{10,14}			\nonumber \\
	&+&a_{11,12}+a_{11,15}+ a_{12,16}+a_{13,16}								\nonumber \\
	&+&a_{14,16}+a_{15,16} + \text{h.c.}, 	\label{eq:lambda}								\\
	\xi &=& a_{1,6}+a_{1,7}+a_{1,8}+a_{1,9}+a_{1,10}+a_{1,11}				\nonumber \\
	&+&a_{2,12}+a_{2,13}+a_{2,14}+a_{3,12}+a_{3,13}+a_{3,15} \nonumber\\
	&+&a_{4,12}+a_{4,14}+a_{4,15}+a_{5,13}+a_{5,14}+a_{5,15}+a_{6,16}\nonumber \\
	&+&a_{7,16}+a_{8,16}+a_{9,16}+a_{10,16}+a_{11,16} + \text{h.c.}.\label{eq:xi}
\end{eqnarray}

% Supplementary bibliography
%	CHANGE THIS TO YOUR OWN BIB FILE
%\putbib[C:/Users/seval/Dropbox/OQuL/OQuLBib/OQuL]
\putbib[topo4thermo-20260717-arXiv]
\end{bibunit}

%^^^^^^^^^^^^^^^^^^^^^^^^^^^^^^^^^^^^^^^^^^^^^^^^^^^^^^^^^^^^^^^^^^^^^^^^^^^^^^^^^^^^^^^^^^^^^^^
%vvvvvvvvvvvvvvvvvvvvvvvvvvvvvvvvvvvvvvvvvvvvvvvvvvvvvvvvvvvvvvvvvvvvvvvvvvvvvvvvvvvvvvvvvvvvvvv
%^^^^^^^^^^^^^^^^^^^^^^^^^^^^^^^^^^^^^^^^^^^^^^^^^^^^^^^^^^^^^^^^^^^^^^^^^^^^^^^^^^^^^^^^^^^^^^^
%vvvvvvvvvvvvvvvvvvvvvvvvvvvvvvvvvvvvvvvvvvvvvvvvvvvvvvvvvvvvvvvvvvvvvvvvvvvvvvvvvvvvvvvvvvvvvvv
%^^^^^^^^^^^^^^^^^^^^^^^^^^^^^^^^^^^^^^^^^^^^^^^^^^^^^^^^^^^^^^^^^^^^^^^^^^^^^^^^^^^^^^^^^^^^^^^
%vvvvvvvvvvvvvvvvvvvvvvvvvvvvvvvvvvvvvvvvvvvvvvvvvvvvvvvvvvvvvvvvvvvvvvvvvvvvvvvvvvvvvvvvvvvvvvv
\end{document}